\documentclass[aps,prl,twocolumn,a4paper,floatfix]{revtex4}

\usepackage{graphicx}
\usepackage{color}
\usepackage{amsmath}
\usepackage{amssymb}
\usepackage{bm}
\usepackage{tikz}

\graphicspath{{./figureExports/}}
\DeclareGraphicsExtensions{.pdf,.jpg}

\usepackage[english]{babel}

\usepackage{lipsum}

\begin{document}
  
  \title{Diffraction catastrophes and semiclassical quantum mechanics for Veselago lensing in graphene}
  
  \author{K. J. A. Reijnders, M. I. Katsnelson}
  \affiliation{Radboud University, Institute for Molecules and Materials,
  Heyendaalseweg 135, 6525 AJ Nijmegen, The Netherlands}

  \email{K.Reijnders@science.ru.nl}
  
  \date{\today}

  \begin{abstract}
    We study the effect of trigonal warping on the focussing of electrons by \emph{n-p} junctions in graphene. We find that perfect focussing, which was predicted for massless Dirac fermions, is only preserved for one specific sample orientation. In the general case, trigonal warping leads to the formation of cusp caustics, with a different position of the focus for graphene's two valleys. We develop a semiclassical theory to compute these positions and find very good agreement with tight-binding simulations.
    Considering the transmission as a function of potential strength, we find that trigonal warping splits the single Dirac peak into two distinct peaks, leading to valley polarization.
    We obtain the transmission curves from tight-binding simulations and find that they are in very good agreement with the results of a billiard model that incorporates trigonal warping.
    Furthermore, the positions of the transmission maxima and the scaling of the peak width are accurately predicted by our semiclassical theory.
    Our semiclassical analysis can easily be carried over to other Dirac materials, which generally have different Fermi surface distortions.
  \end{abstract}

  \maketitle

  Veselago lenses~\cite{Veselago68} are special types of lenses, which are made of materials with a negative refractive index. 
  Such lenses can overcome the diffraction limit~\cite{Pendry00} and can nowadays be realized in metamaterials~\cite{Smith00,Houck03,Grbic04}, chiral metamaterials~\cite{Tretyakov03,Pendry04,Zhang09,Xiong10} and photonic crystals~\cite{Parimi04,Cubukcu03}.
  An electronic analog of a Veselago lens can be created using \emph{n-p} junctions, with the classical trajectories of the charge carriers playing the role of the rays in geometrical optics. Such junctions focus electrons, because the group velocity for holes is in the direction opposite to their phase velocity, whereas the two velocities are in the same direction for electrons.
  However, in conventional semiconductors, such interfaces are unsuitable because of their high reflectivity, owing to the presence of a depletion region.
  
  Cheianov et al.~\cite{Cheianov07} realized that graphene does not have this drawback. It has zero bandgap, as the valence and conduction bands touch at two non-equivalent corners of the Brillouin zone, known as $K$ and $K'$. The low-energy charge carriers are ballistic over large distances and follow the Dirac equation~\cite{Wallace47,McClure57,Slonczewski58,Semenoff84,CastroNeto09,Katsnelson13}. This gives rise to Klein tunneling: normally incident electrons are transmitted with unit probability~\cite{Klein29,Katsnelson06,Cheianov06,Shytov08,Tudorovskiy12,Reijnders13,Young09,Stander09,Chen16}, which makes the interface exceptionally transparent. 
  Recently, Veselago lensing in graphene was experimentally observed~\cite{Lee15,Chen16}.

  Theoretical studies have used the Dirac equation to investigate focussing by flat~\cite{Cheianov07,Reijnders17} and circular~\cite{Cserti07,Peterfalvi10,Wu14} junctions and in zigzag nanoribbons~\cite{Choi14}. It was shown that when the electron and hole charge carrier densities are equal, a flat interface is able to focus all trajectories into a single point~\cite{Cheianov07}. According to catastrophe theory~\cite{Berry80,Poston78,Arnold82,Arnold75},
  such a situation is exceptional: any perturbation of the Hamiltonian will ruin this ideal focus. 
  Indeed, in tight-binding simulations of an \emph{n-p} junction~\cite{Milovanovic15}, where transmission was studied as a function of potential strength, a direction-dependent broadening of the transmission peak was observed. However, the precise origin of this broadening was not clarified.
  
  An important correction to the Dirac Hamiltonian is the second-order term in the expansion of graphene's tight-binding Hamiltonian, known as the trigonal warping term~\cite{Ajiki96,Ando98,Katsnelson13}. 
  When adding this term, the Hamiltonian becomes dependent on the crystallographic direction of the sample and becomes different for the two valleys, leading to different classical trajectories.
  Using this principle, a valley beam splitter based on an \emph{n-p-n'} junction was devised~\cite{Garcia08}, in which the trajectories in the $K$- and $K'$-valleys are deflected in different angular directions. Creating valley polarization~\cite{Rycerz07,Xiao07,Gorbachev14} is important for graphene valleytronics applications, where the valley index is used to encode information in a way similar to spintronics.

  \begin{figure*}[!htb]
  \includegraphics{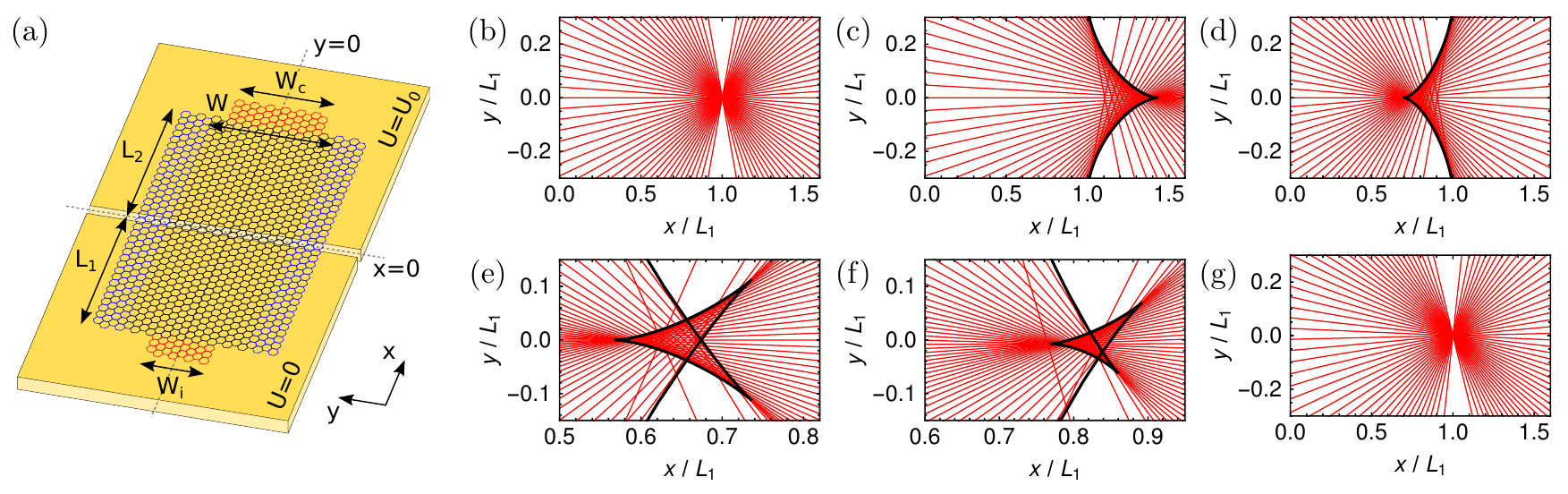}
  \caption{(a) Simulation setup with an injector and collector lead (red) and drain leads on each side (blue). (b) Classical trajectories for the massless Dirac Hamiltonian at $U_0=2E$. (c)--(g) Classical trajectories (red) and caustics (black) for the Hamiltonian including trigonal warping. Unless otherwise indicated, $E=0.4$ eV. (c) $K$-valley, $U_0=0.8$~eV, $\theta=0$, (d) $K'$-valley, $U_0=0.8$~eV, $\theta=0$, (e) Section of the butterfly caustic. $K'$-valley, $E=0.6$~eV, $U_0=1.18$~eV, $\theta=0$, (f) $K'$-valley, $U_0=0.795$~eV, $\theta = \pi/12$, (g) $U_0=0.8$~eV, $\theta=\pi/6$.}
  \label{fig:setup-trajectories}
  \end{figure*}
  
  In this paper, we develop a complete semiclassical theory of Veselago lensing by an \emph{n-p} junction in the presence of trigonal warping. We show that the ideal focus generally disappears and analyze the different diffraction catastrophes, known as caustics~\cite{Berry80,Poston78,Arnold82,Arnold75}, that are formed. We show that electrons from the $K$- and $K'$-valleys are generally focussed at different positions, even at moderate energies.
  We obtain these positions using semiclassical methods~\cite{Connor81b,Reijnders17,Dobrokhotov14} based on the Pearcey function~\cite{Pearcey46,Connor81a,Connor82} and find very good agreement with tight-binding simulations using the Kwant code~\cite{Groth14}.
  We also show that an initial sublattice polarization leads to tilting of the focus.
  Furthermore, we obtain the transmission as a function of the hole carrier density from both tight-binding simulations~\cite{Groth14} and a semiclassical billiard model~\cite{Beenakker89,Milovanovic13} that incorporates trigonal warping. We find that trigonal warping explains the previously observed peak broadening~\cite{Milovanovic15} and that our semiclassical theory accurately predicts the positions of the transmission maxima and the scaling of the peak widths.
  Since the transmission maximum occurs at a different potential for the two valleys, we find that a graphene \emph{n-p} junction can act as a valley filter and that one can manipulate the polarization by changing the potential strength.
  This is similar to the situation in chiral metamaterials~\cite{Tretyakov03,Pendry04,Zhang09,Xiong10}, where one can select a certain polarization by tuning the frequency of the incident light.

  Trigonal warping adds terms that are quadratic in momentum to the linear term in the massless Dirac Hamiltonian~\cite{Ajiki96,Ando98,Katsnelson13}. We make all terms in the Hamiltonian dimensionless~\cite{Reijnders13} by scaling energies with the electron energy $E$ and lengths with $L_1$, the distance from the source to the junction, see Fig.~\ref{fig:setup-trajectories}(a). Using first-order perturbation theory, we obtain the classical Hamiltonian in dimensionless variables~\cite{Ajiki96,Ando98,Katsnelson13,Berlyand87,Belov06}:
  \begin{equation}
    H_\alpha^\pm = \pm \Big( | \mathbf{p} | + \alpha \mu | \mathbf{p} |^2 \cos\big[3 (\phi_{\mathbf{p}}+\theta )\big] \Big) + U(\mathbf{x}),
    \label{eq:H-class-trigwarp-dimless-main}
  \end{equation}
  where $\mu=E/6t$, with $t$ the nearest-neighbor hopping, indicates the relative importance of the quadratic term and $\alpha$ equals +$1$ ($-1$) for valley $K'$ ($K$). The angle $\phi_{\mathbf{p}}=\arctan(p_y/p_x)$ and $\theta$ determines the orientation of the sample, with $\theta=0$ corresponding to zigzag edges along the $x$-direction. The dimensionless semiclassical parameter $h=3ta_{CC}/2EL_1$, with $a_{CC}$ the carbon-carbon distance.

  \begin{figure*}[!htb]
  \includegraphics{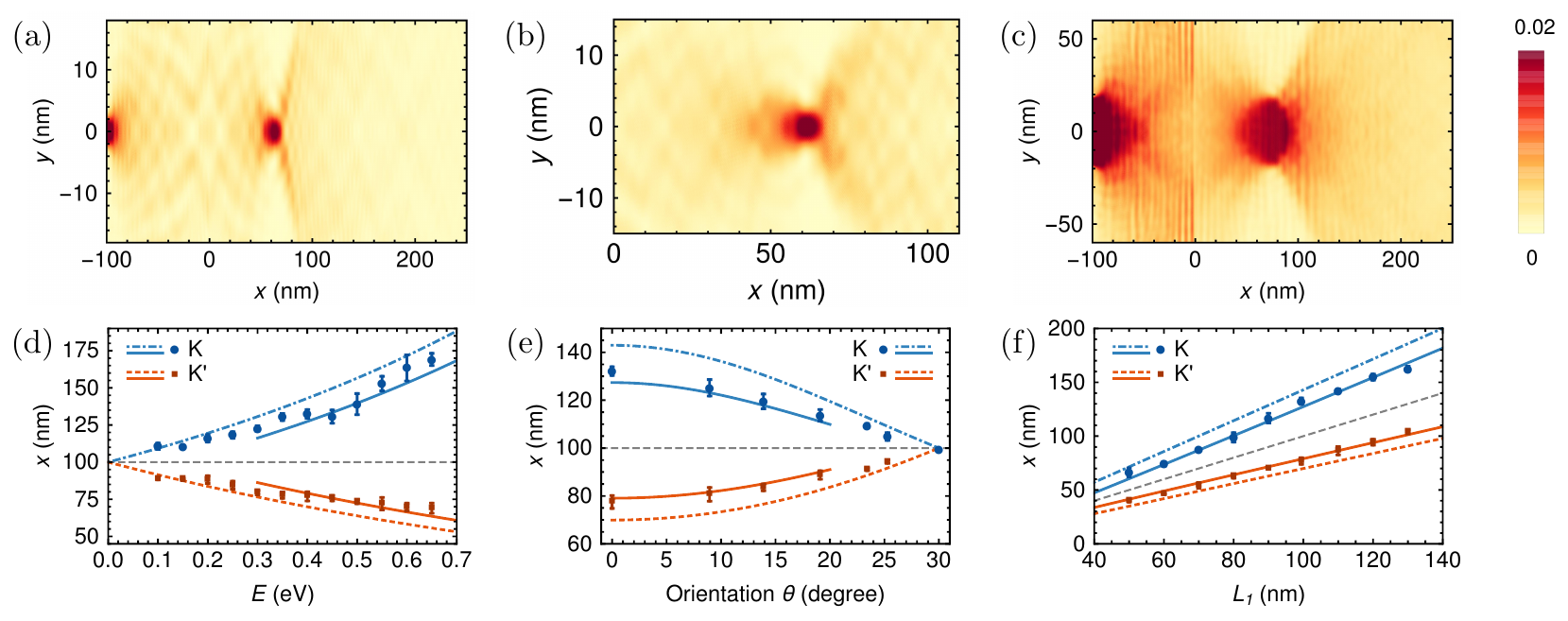}
  \caption{(a)--(c) Results of the tight-binding simulations with $L_1=100$~nm. The density $|\Psi_{\text{av},\alpha}|^2$ is averaged over sublattices and summed over lead modes in valley $\alpha$. (a) $K'$-valley, $E=0.6$~eV, $U_0=2E$, $W_i=7.5$~nm;
  (b) $K'$-valley, $E=0.6$~eV, $U_0=1.18$~eV, $W_i=7.5$~nm; cf. the classical trajectories in Fig.~\ref{fig:setup-trajectories}(e); (c) $K'$-valley, $E=0.4$~eV, $U_0=2E$, $W_i=40$~nm. (d)--(f) Position, on the $x$-axis, of the caustic (dashed and dashed-dotted lines), semiclassical maximum (solid lines) and simulated maximum (symbols) for varying energy $E$, sample orientation $\theta$ and $L_1$. The dashed grey lines indicate the Dirac result. The parameters equal, (e, f) $E=0.4$~eV; (d, f) $\theta=0$; (d, e) $L_1=100$~nm; (d, f) $W_i=40$~nm; (e) $W_i=50$~nm. In all cases $U_0=2E$.}
  \label{fig:real-space-wavefunction-maxima}
  \end{figure*}
  
  We consider a sharp, one-dimensional junction: $U(\mathbf{x})=U_0\Theta(x)$, with $\Theta(x)$ the Heaviside step function. Then $p_y$ is conserved and the classical action for an electron with energy $E$ emitted from $(-L_1,0)$ equals~\cite{Cheianov07,Reijnders17}
  \begin{equation}
    S_{np}(x,y,p_y) = L_1 \, p_{x,e,\alpha}(p_y) + x \, p_{x,h,\alpha}(p_y) + y \, p_y ,
    \label{eq:action-dimless}
  \end{equation}
  with $p_{x,e,\alpha}$ ($p_{x,h,\alpha}$) the longitudinal momentum in electron (hole) region, i.e. $H_\alpha^+(p_{x,e,\alpha}(p_y),p_y)=E$.
  The classical trajectories, i.e. the points where $\partial S_{np}/\partial p_y=0$, and the caustics, where $\partial^2S_{np}/\partial p_y^2$ vanishes as well, are plotted in Fig.~\ref{fig:setup-trajectories} for various parameters. The Dirac Hamiltonian, $\mu=0$ in Eq.~(\ref{eq:H-class-trigwarp-dimless-main}), is symmetric in $p_x$ and $p_y$. When $U_0=2E$, this leads to an ideal focus~\cite{Cheianov07}, at which all derivatives of $S_{np}$ with respect to $p_y$ vanish~\cite{Reijnders17}.
  This generally changes when we include trigonal warping. For $\theta=0$, the symmetry in $p_y$ is preserved, leading to reflection symmetry in the $x$-axis. However, the symmetry in $p_x$ is broken and we obtain cusp caustics, shown in Fig.~\ref{fig:setup-trajectories}(c) and (d). Because we also break the symmetry between the valleys $K$ and $K'$, their cusp points, at which $\partial^4S_{np}/\partial p_y^4\neq0$, are at different positions on the $x$-axis. Tuning the potential, we obtain sections of the butterfly caustic, see Fig.~\ref{fig:setup-trajectories}(e), and eventually pass through the butterfly singularity, at which $\partial^4S_{np}/\partial p_y^4=0$, but $\partial^6S_{np}/\partial p_y^6\neq0$~\cite{Reijnders17,Poston78}. For generic $\theta$, Fig.~\ref{fig:setup-trajectories}(f), both symmetries are broken and the cusp point is no longer on the $x$-axis. Only when $\theta=\pi/6$ (armchair edges), the symmetry in $p_x$ is restored and we recover an ideal focus at $U_0=2E$.

  From here on, we consider zigzag edges along the $x$-direction ($\theta=0$), unless otherwise indicated, as they illustrate the generic situation. We obtain an expression for $x_{\text{cusp},\alpha}$ by solving $\partial^2S_{np}/\partial p_y^2=0$ for $x$, and setting $p_y=0$~\cite{Reijnders17}.
  Expanding the result up to second order in $\alpha \mu$, we find, in units with dimensions
  \begin{equation}
    x_{\text{cusp},\alpha} = L_1 \frac{U_0-E}{E} \left( 1 - \alpha \frac{4 U_0}{3 t} \right)
    \stackrel{U_0=2 E}{=\mathrel{\mkern-3mu}=\mathrel{\mkern-3mu}=\mathrel{\mkern-3mu}=} 
    L_1 - \frac{8 \alpha E}{3 t} L_1 .
    \label{eq:x-cusp-trigwarp}
  \end{equation}
  Hence, the cusp point for the valley $K$ ($K'$) is always to the right (left) of the focus for the Dirac Hamiltonian, given by the first term. Although the above expansion is only sufficient for low energies, it clearly indicates that the effect is sizeable.

  We study these effects by performing tight-binding simulations with Kwant~\cite{Groth14}. We use the setup of Ref.~\cite{Milovanovic15}, shown in Fig.~\ref{fig:setup-trajectories}(a), where current enters the sample through an injector lead and is able to exit through a collector lead and through drain leads on each side~\cite{Supplementary}. Considering large $L_2$ and $W_c=W$, we compute the sample wavefunction.
  Figure~\ref{fig:real-space-wavefunction-maxima}(a)--(c) shows the resulting density, averaged over sublattices and summed over lead modes~\cite{Reijnders17,Brey06}. For narrow leads, i.e. large $h_{\text{lead}}=\hbar v_F/EW_i$, we observe cusp caustics in the valleys $K$ and $K'$, see Fig.~\ref{fig:real-space-wavefunction-maxima}(a). By adjusting $E$ and $U_0$, we also observe sections of the butterfly caustic, see Fig.~\ref{fig:real-space-wavefunction-maxima}(b).
  Figure~\ref{fig:real-space-wavefunction-maxima}(c) shows that we obtain a bright focussing spot for wide leads (small $h_{\text{lead}}$), as predicted in Ref.~\cite{Reijnders17}.
  Subsequently, we fit a Gaussian to a subset of the points on the $x$-axis. Averaging over subsets, we obtain the position of the maximum and an error estimate.
  
  Within the semiclassical approximation, the wavefunction near a cusp can be constructed in terms of the Pearcey function~\cite{Connor81b,Reijnders17,Dobrokhotov14,Supplementary}.
  We find the parameters for this function by expanding the action~(\ref{eq:action-dimless}) up to fourth order in $p_y$.
  Figure~\ref{fig:real-space-wavefunction-maxima}(d)--(f) shows the predictions $x_{\text{max},\alpha}$, obtained from maximizing this wavefunction, as a function of electron energy, sample orientation and $L_1$. They are in very good agreement with the positions extracted from the simulations. Note that the approximation fails for low energies, or near $\theta=\pi/6$, as we are too close to the ideal focus, at which $\partial^4S_{np}/\partial p_y^4=0$. However, in these cases, $x_{\text{cusp},\alpha}$ provides a good estimate.

  \begin{figure*}[!htb]
  \includegraphics{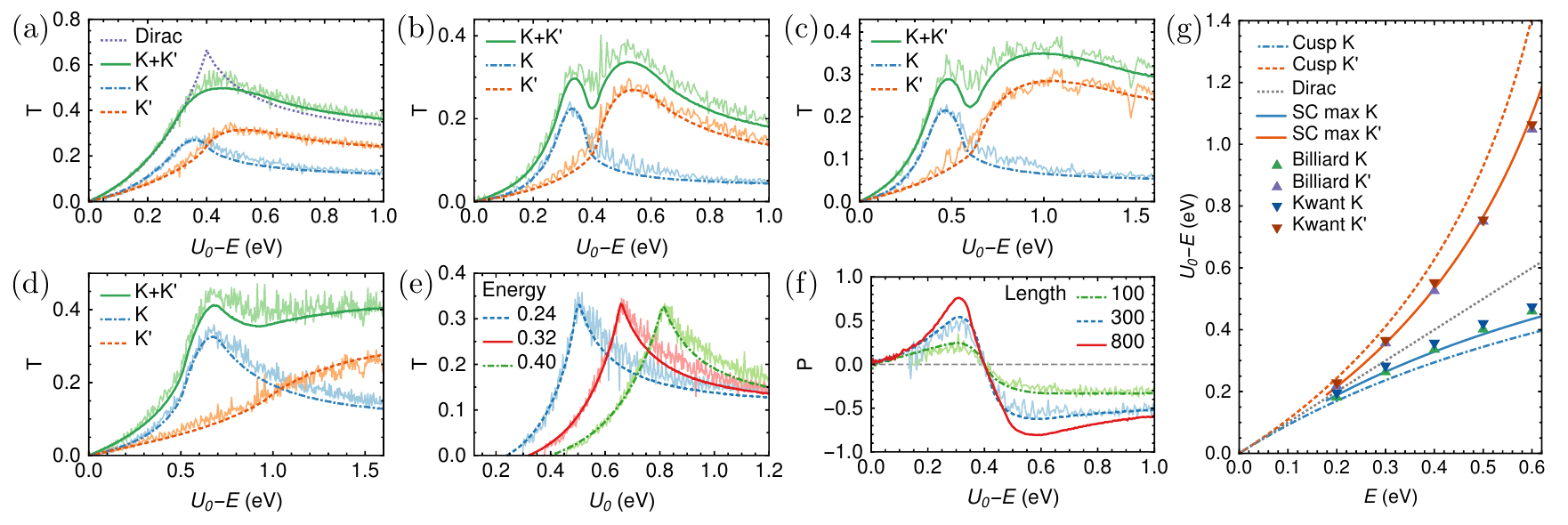}
  \caption{(a)--(e) Intravalley transmission as a function of potential strength obtained from Kwant (light colors, wavy lines) and a billiard model (darker colors, smooth lines). (a) $E=0.4$~eV, $L_1=L_2=100$~nm; (b) $E=0.4$~eV, $L_1=L_2=300$~nm; (c) $E=0.6$~eV, $L_1=L_2=200$~nm; 
    (d) $E=0.4$~eV, $L_1=100$~nm, $L_2=300$~nm. 
    (e) $K$-valley, $L_1=100$~nm, $L_2=140$~nm for various energies (in eV). For $E=0.24$~eV, $\partial^4 S_{np}/\partial p_y^4$ vanishes at $U_0=0.5$~eV and $x=L_2$.
    (f) Valley polarization $P=(T_K-T_{K'})/(T_K+T_{K'})$ as a function of potential for $E=0.4$~eV and various $L_1=L_2$ (in nm).
    (g) Maximum of the simulated intensity as a function of energy for $L_1=L_2=200$~nm. We also plot the potentials at which $x_{\text{cusp},\alpha}$ and the semiclassical (SC) $x_{\text{max},\alpha}$ equal $L_2$. For all figures $W_i=W_c=50$~nm.
  }
  \label{fig:T-U-dep}
  \end{figure*}

  \begin{figure*}[!htb]
  \includegraphics{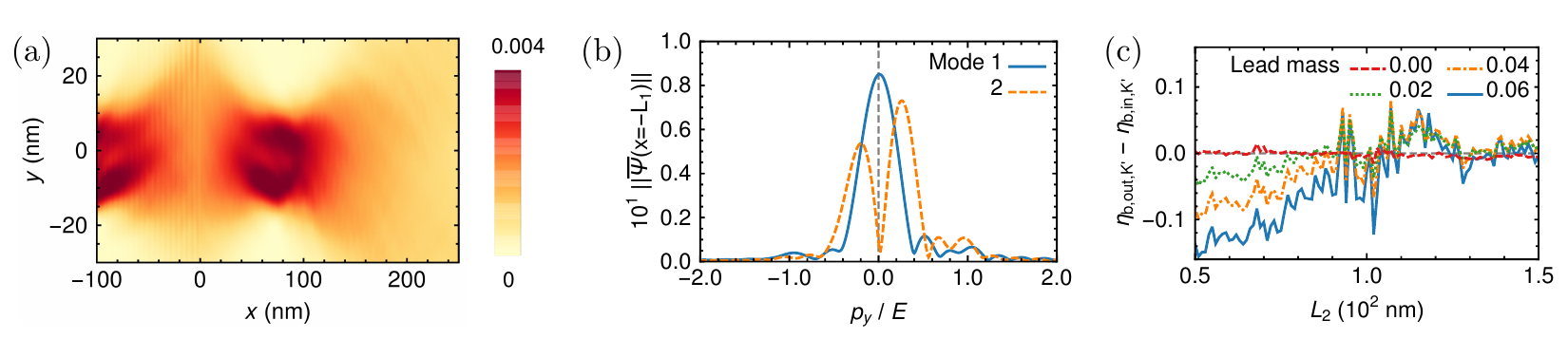}
  \caption{(a) Density $|\Psi_{\text{av},K'}|^2$ from tight-binding simulations for $E=0.4$~eV, $m_{\text{lead}}=0.375$~eV, $W_i=40$~nm.
    (b) Fourier transform (for each lead mode) of the wavefunction from (a) at $x=-L_1$. (c) Symmetry breaking as a function of $L_2$ for various  $m_{\text{lead}}$ (in eV). $K'$-valley, $E=0.1$~eV, $W_i=50$~nm.
    For all figures $L_1=100$~nm and $U_0=2E$.
  }
  \label{fig:mass-influence-main}
  \end{figure*}
  
  Experimentally, one typically measures the transmission $T$ as a function of the potential strength. We therefore fix $L_1$ and $L_2$, set $W_c=W_i$ and compute the intravalley scattering for varying $U_0$ using Kwant.
  We compare these results with a semiclassical billiard model~\cite{Beenakker89,Milovanovic13} that incorporates trigonal warping in the trajectories and has uniformly distributed initial positions and transversal momenta.
  Figure~\ref{fig:T-U-dep} shows that the agreement between the two methods is very good. Note that the tight-binding transmission is slightly higher than the billiard result on the right (left) of the peak for valley $K$ ($K'$), most likely because of interference. Higher order effects and intervalley scattering only mildly influence the results.
  Figure~\ref{fig:T-U-dep}(e) shows that, for small $h_{\text{lead}}$, passing through the butterfly singularity does not significantly enhance the transmission.

  Valley splitting is observed at all simulated energies and the total transmission peak is broadened with respect to the result of a Dirac billiard model, which explains earlier observations~\cite{Milovanovic15}. The transmission maximum for the $K$-valley is always at a lower potential. Setting $x_{\text{cusp},\alpha}=L_2$ in Eq.~(\ref{eq:x-cusp-trigwarp}) and solving for $U_0$, we obtain, up to $\mathcal{O}(\mu^2)$, in units with dimensions,
  \begin{equation}
    U_{0,\text{cusp},\alpha} \!= 
    E \frac{L_1+L_2}{L_1} \left( 1 + \alpha \frac{4 E}{3 t} \frac{L_2}{L_1} \right)
    \stackrel{L_1=L_2}{=\mathrel{\mkern-4mu}=\mathrel{\mkern-4mu}=\mathrel{\mkern-4mu}=}
    2E+ \frac{8 \alpha E^2}{3 t}.
  \end{equation}
  We also obtain predictions $U_{0,\text{max},\alpha}$ for the positions of the maxima from our semiclassical theory. Solving $x_{\text{max},\alpha}=L_2$, we obtain values that are in excellent agreement with the simulated peak positions, see Fig.~\ref{fig:T-U-dep}(g).
  As $\partial^4S_{np}/\partial p_y^4$ is lower at the $K$-peak, our theory also correctly predicts that it has lower maximal transmission.
  
  The valley polarization can be increased by going to higher energies, which increases the peak splitting. Alternatively, one can increase $L_2$, see Fig.~\ref{fig:T-U-dep}(d). Finally, one can also increase both $L_1$ and $L_2$, see Fig.~\ref{fig:T-U-dep}(f), which decreases the peak width $\Delta U_{0}$.
  We obtain a semiclassical prediction for the scaling behavior of $\Delta U_{0}$ by performing a Taylor expansion of the equation $x_{\text{max},\alpha}=L_2$ around $U_{0,\text{cusp},\alpha}$. For $L_1=L_2=L$, we find that $\Delta U_0\propto L^{-1/2}$ in leading order, which is in perfect agreement with the billiard results.

  Finally, we briefly discuss symmetry breaking due to an initial sublattice polarization~\cite{Reijnders17}. 
  When we add a constant mass $m_{\text{lead}}$ in the lead, we observe tilting of the main focus, see Fig.~\ref{fig:mass-influence-main}. 
  Inspecting the Fourier transform, we observe that it is not symmetric in $p_y$, verifying earlier predictions~\cite{Reijnders17}.
  To certify that this effect can be attributed to the Dirac character of the electrons, and not to trigonal warping, we extract the wavefunction at $x=-L_1$ from Kwant and compute its evolution using the continuum Dirac equations~\cite{Reijnders17}.
  This reproduces the tilting, but changes the position of the focus, confirming our hypothesis.
  
  The symmetry breaking can be quantified by setting $W_c=W$ and subsequently splitting the collector lead at $y=0$. We then extract the relative transmission $\eta_{b,\text{out},\alpha}$ through the bottom ($y<0$) collector lead and subtract the relative intensity $\eta_{b,\text{in},\alpha}$ located on the sites with negative $y$ for the incoming modes. Figure~\ref{fig:mass-influence-main}(c) shows that the amount of symmetry breaking strongly depends on $L_2$, as expected for a tilted focus. 
  In agreement with predictions made for the Green's function~\cite{Reijnders17}, it increases with increasing mass, decreases with increasing energy, and changes sign when we change the sign of the mass. Simulating transmission as a function of potential strength, initial polarization seems to slightly decrease the valley polarization.
  
  In short, we have developed a complete semiclassical theory of Veselago lensing in the presence of trigonal warping. We have observed both cusp and butterfly singularities, which provide an interesting relation between the physics of graphene and catastrophe theory~\cite{Berry80,Poston78,Arnold82,Arnold75}. Our theory shows excellent agreement with tight-binding simulations and explains previously observed peak broadening~\cite{Milovanovic15}.
  We believe that our predictions could be experimentally verified, using, for instance, a transverse magnetic focussing setup~\cite{Chen16}. Although this would require rather high energies, these have been experimentally realized~\cite{Wehling08,Nair13,Lee16,Khademi16}.
  We remark that a smooth junction will lead to additional peak broadening~\cite{Phong16,Milovanovic15,Chen16}, but this makes a theoretical study much more complicated~\cite{Reijnders13}.
  
  We emphasize that our analysis of the disintegration of the ideal focus does not fundamentally depend on the fact that the Fermi surface distortion is trigonal: any distortion that breaks the symmetry in $p_x$ will lead to the formation of cusp caustics.
  Therefore, our semiclassical analysis can easily be carried over to other Dirac materials, such as topological insulators~\cite{Hasan10,Qi11}. Since these generally exhibit stronger band bending, the deviations from the Dirac behavior will be much stronger and therefore visible at lower energies.

  \begin{acknowledgments}
    We are grateful to Misha Titov and Erik van Loon for helpful discussions.
    The authors acknowledge support from the ERC Advanced Grant 338957 FEMTO/NANO and from the NWO via the Spinoza Prize.
  \end{acknowledgments}

  \clearpage

  \onecolumngrid
  
  \setcounter{table}{0}
  \setcounter{section}{0}
  \setcounter{figure}{0}
  \setcounter{equation}{0}
  \renewcommand{\thesection}{S\arabic{section}}
  \renewcommand{\thetable}{S\arabic{table}}
  \renewcommand{\thefigure}{S\arabic{figure}}
  \renewcommand{\theequation}{S\arabic{equation}}

  \section{Supplemental material}

  \section{Detailed derivations}

  For the convenience of the reader, this section contains the details of the derivations that have been discussed in the main text. The semiclassical theory that we develop is based on methods that have been well-established in the literature. For the Dirac Hamiltonian, a similar theory was presented in Ref.~\cite{Reijnders17}.

  \subsection{Hamiltonian with trigonal warping}
  
  In reciprocal space, the full tight-binding Hamiltonian for graphene is given by (see e.g. Ref.~\cite{Katsnelson13})
  \begin{equation}
    H_{\text{full}} = 
    \begin{pmatrix}
      0 & -t u(\mathbf{k}) \\
      -t u(\mathbf{k})^* & 0 
    \end{pmatrix}
    , \qquad \text{ where } u(\mathbf{k}) = \sum_{\bm{\delta}_i} e^{i \mathbf{k} \cdot \bm{\delta}_i} ,
  \label{eq:H-matrix-full}
  \end{equation}
  where the nearest-neighbor vectors $\bm{\delta}_i$ are given by
  \begin{equation}
    \bm{\delta}_1 = a_{CC} \left(0, 1 \right) , \quad \bm{\delta}_2 = \frac{a_{CC}}{2} \left(\sqrt{3}, -1 \right), \quad \bm{\delta}_3 = \frac{a_{CC}}{2} \left(-\sqrt{3}, -1 \right) .
  \end{equation}
  Furthermore, we choose our lattice vectors $\mathbf{v}_i$ as
  \begin{equation}
    \mathbf{v}_1 = a_{CC} \left(\sqrt{3}, 0 \right), \quad \mathbf{v}_2 = \frac{a_{CC}}{2} \left(\sqrt{3}, 3 \right) .
    \label{eq:lattice-vecs}
  \end{equation}
  The electron and hole bands touch at two non-equivalent Dirac points in the Brillouin zone, named $K$ and $K'$:
  \begin{equation}
    \mathbf{K} = \frac{2 \pi}{3 \sqrt{3} a_{CC}} \left( -1, \sqrt{3} \right), \qquad \mathbf{K}' = \frac{2 \pi}{3 \sqrt{3} a_{CC}} \left(1, \sqrt{3} \right) .
  \end{equation}
  Expanding the Hamiltonian~(\ref{eq:H-matrix-full}) up to second order in $\mathbf{k}$ around the $K$-point, we obtain~\cite{Ajiki96,Ando98,Katsnelson13}
  \begin{equation}
    H_K =
    \begin{pmatrix} 
      0 & \frac{3}{2} t a_{CC} (k_x - i k_y) - \frac{3}{8} t a_{CC}^2 (k_x + i k_y)^2 \\
      \frac{3}{2} t a_{CC} (k_x + i k_y) - \frac{3}{8} t a_{CC}^2 (k_x - i k_y)^2 & 0
    \end{pmatrix} .
    \label{eq:H-matrix-trigwarp-K}
  \end{equation}
  Keeping only the linear term, we obtain the Dirac Hamiltonian. The quadratic term is the trigonal warping correction.
  Expanding the Hamiltonian~(\ref{eq:H-matrix-full}) around the $K'$-point, we find
  \begin{equation}
    H_{K'} =
    \begin{pmatrix} 
      0 & \frac{3}{2} t a_{CC} (k_x + i k_y) + \frac{3}{8} t a_{CC}^2 (k_x - i k_y)^2 \\
      \frac{3}{2} t a_{CC} (k_x - i k_y) + \frac{3}{8} t a_{CC}^2 (k_x + i k_y)^2 & 0
    \end{pmatrix} .
    \label{eq:H-matrix-trigwarp-Kpr} 
  \end{equation}
  Considering the trigonal warping term as a perturbation to the Dirac Hamiltonian, we compute the first order correction to the dispersion relation using perturbation theory and obtain~\cite{Ajiki96,Ando98,Katsnelson13}
  \begin{equation}
    E_\alpha^\pm = \pm \left( \frac{3}{2} t a_{CC} | \mathbf{k} | + \alpha \frac{3}{8} t a_{CC}^2 | \mathbf{k} |^2 \cos(3 \phi_{\mathbf{k}}) \right) ,
    \label{eq:E-class-trigwarp}
  \end{equation}
  where the valley index $\alpha$ equals $-1$ for the $K$-valley and $+1$ for the $K'$-valley. Furthermore, $\phi_{\mathbf{k}}$ is the angle in $k$-space, defined by $\tan \phi_{\mathbf{k}} = k_y/k_x$.
  Let us now consider an electron with energy $E$ and let $L_1$ be the typical length scale of the problem, in our case the distance from the injector lead to the junction interface. We then define the dimensionless variables $\tilde{E}_\alpha^\pm =  E_\alpha^\pm/E$, $\tilde{\mathbf{x}}=\mathbf{x}/L_1$, $\tilde{\mathbf{k}}=\mathbf{k} L_1$ and the dimensionless semiclassical parameter $h=3 t a_{CC}/(2 E L_1)$. The dimensionless coefficient $\mu = E/(6 t)$ controls the relative importance of the second term and we require $\mu^2$ to be small in order for our perturbative expansion to be valid. Introducing $\tilde{\mathbf{p}}=h \tilde{\mathbf{k}}$ and omitting the tildes, we obtain 
  \begin{equation}
    E_\alpha^\pm = \pm \left( | \mathbf{p} | + \alpha \mu | \mathbf{p} |^2 \cos(3 \phi_{\mathbf{p}}) \right) .
    \label{eq:E-class-trigwarp-dimless-noAngle}
  \end{equation}
  We remark that $\phi_{\mathbf{k}}=\phi_{\mathbf{\tilde{k}}}=\phi_{\mathbf{\tilde{p}}}$. We will, from here on, omit the tildes and exclusively work with these dimensionless variables, unless otherwise indicated. 
  
  Let us now consider a passive rotation by an angle $\theta$, which means that we rotate the coordinate system over an angle $\theta$. Then the nearest-neighbor vectors $\bm{\delta}_i$ should be multiplied (from the left) by the matrix of a rotation over an angle $-\theta$. Because of this, the expressions for the Dirac points should be multiplied by the same factor. In turn, this leads to an additional factor $e^{3i\theta}$ in front of the quadratic term in the expansion of $u(\mathbf{k})$ around the $K$-point, and to an additional factor $e^{-3i\theta}$ in its expansion around the $K'$-point. Because of these additional factors, our dimensionless expression for the energy then becomes
  \begin{equation}
    E_\alpha^\pm = \pm \Big( | \mathbf{p} | + \alpha \mu | \mathbf{p} |^2 \cos\big[3 (\phi_{\mathbf{p}}+\theta)\big] \Big) .
    \label{eq:E-class-trigwarp-dimless}
  \end{equation}

  Using standard semiclassical arguments~\cite{Berlyand87,Belov06}, one can then show that the classical Hamiltonian functions for electrons and holes are given by
  \begin{equation}
    H_\alpha^\pm = \pm \Big( | \mathbf{p} | + \alpha \mu | \mathbf{p} |^2 \cos\big[3 (\phi_{\mathbf{p}}+\theta )\big] \Big) + U(\mathbf{x}),
    \label{eq:H-class-trigwarp-dimless}
  \end{equation}
  It is important to note that this Hamiltonian is no longer invariant with respect to the orientation of the sample, as indicated by the presence of the orientation angle $\theta$. In our choice of coordinates, $\theta=0$ corresponds to zigzag edges along the $x$-axis and $\theta=\pi/6$ corresponds to armchair edges along the $x$-axis. Furthermore, we remark that the lattice is invariant under a $\pi/3$ rotation, but that such a rotation interchanges the sublattices $A$ and $B$ and the valleys $K$ and $K'$.

  \subsection{Caustics}
  
  We consider an \emph{n-p} junction with the one-dimensional potential
  \begin{equation}
    U(\mathbf{x}) = U_0 \Theta(x) ,
  \end{equation}
  where $\Theta(x)$ is the Heaviside step function. As the potential is one-dimensional, the transversal momentum $p_y$ is a conserved quantity.
  For $x<0$, the charge carriers are electrons with longitudinal momentum $p_{x,e,\alpha}$ that satisfies $E_\alpha^+(p_{x,e,\alpha}, p_y) = E$. For $x>0$, we have holes with longitudinal momentum $p_{x,h,\alpha}$ that satisfies $E_\alpha^-(p_{x,h,\alpha}, p_y) +U_0 = E$.
  
  Let us study electrons that are emitted by a point source at $(-L_1,0)$ and are incident on this potential. It easily follows that the classical action is given by~\cite{Cheianov07,Reijnders17}
  \begin{equation}
    S_{np}(x,y,p_y) = L_1 p_{x,e,\alpha}(p_y) + x p_{x,h,\alpha}(p_y) + y p_y .
    \label{eq:action-dimless-app}
  \end{equation}
  The trajectories, which are the stationary points of the action, i.e. the points where $\partial S_{np}/\partial p_y=0$, are given by
  \begin{equation}
    y = - L_1 \frac{\partial p_{x,e,\alpha}}{\partial p_y} - x \frac{\partial p_{x,h,\alpha}}{\partial p_y} .
    \label{eq:trajectories}
  \end{equation}
  The caustic is the set of points $(x_{\text{cst}},y_{\text{cst}})$ where the second derivative $\partial^2 S_{np}/\partial p_y^2$ vanishes as well. Differentiating Eq.~(\ref{eq:trajectories}), we obtain
  \begin{equation}
    x_{\text{cst}} = - L_1 \frac{\partial^2 p_{x,e,\alpha}/\partial p_y^2}{\partial^2 p_{x,h,\alpha}/\partial p_y^2} ,
    \label{eq:x-caustic}
  \end{equation}
  and we can obtain $y_{\text{cst}}$ by entering this value back into Eq.~(\ref{eq:trajectories}). The cusp point $(x_{\text{cusp}},y_{\text{cusp}})$ is the point on the caustic where the third derivative $\partial^3 S_{np}/\partial p_y^3$ is identically zero.
  
  Let us first look at the Dirac case for a moment. Setting $\mu=0$ in Eq.~(\ref{eq:E-class-trigwarp-dimless}), we see that we have rotation symmetry in momentum space. In particular, we have reflection symmetry in $p_x$ and in $p_y$. As $p_y$ is conserved at the barrier interface and the group velocity for holes is antiparallel to their momentum, symmetry in $p_x$ means that $p_{x,h,\alpha}=-p_{x,e,\alpha}$ when $U_0=2E$. This leads to an ideal focus, as the electrons are focussed in the mirror image of the source~\cite{Cheianov07,Reijnders17}. Furthermore, the symmetry $p_y \to -p_y$ ensures that the classical trajectories are also symmetric in the $x$-axis.
  
  When we include trigonal warping, the dispersion relation becomes dependent on the sample orientation $\theta$. For $\theta=0$, which means that we have zigzag edges along the $x$-axis, the symmetry in $p_x$ is broken, but the symmetry in $p_y$ is preserved. Therefore, we no longer have an ideal focus at $U_0=2E$, but the trajectories are still symmetric in the $x$-axis.
  On the other hand, when $\theta=\pi/6$, we do have symmetry in $p_x$, but the symmetry in $p_y$ is broken. Indeed, we immediately see that $\cos(3 [\phi_p + \pi/6]) = -\sin(3 \phi_p)$ and setting $\bar{\mathbf{p}} = (-p_x, p_y)$, we observe that $\sin(3 \phi_{\bar{\mathbf{p}}}) = \sin(3 [\pi - \phi_{\mathbf{p}}]) = \sin(3 \phi_{\mathbf{p}})$. Therefore, we have an ideal focus at $U_0=2E$ when we have armchair edges along the $x$-axis. However, the trajectories are no longer symmetric in the $x$-axis.
  For generic $\theta$, both symmetries are broken and there will be no ideal focus at $U_0=2E$.
  
  Let us now take a closer look at $\theta=0$, as the symmetry in $p_y$ considerably simplifies the calculations.
  Since $E_\alpha^\pm(p_x,-p_y)=E_\alpha^\pm(p_x,p_y)$, we also have $p_{x,e,\alpha}(p_y)=p_{x,e,\alpha}(-p_y)$, and the same identity holds for $p_{x,h,\alpha}$. Therefore, their second derivatives are symmetric with respect to $p_y$, which implies that $x_{\text{cst}}$ is. Hence, we infer from Eq.~(\ref{eq:trajectories}) that the caustic is symmetric in the $x$-axis, which, in particular, means that the cusp points are. As the third derivative $\partial^3 S_{np}/\partial p_y^3$ is antisymmetric in $p_y$, it vanishes at $p_y=0$ and there is always a cusp point on the $x$-axis.
  In what follows, we investigate the position $x_{\text{cusp}}$ of this particular cusp point. 
  We first note that when we expand $S_{np}$ at $(x_{\text{cusp}},0)$ in $p_y$ around $p_y=0$, the terms of odd order vanish.
  Let us now determine the various deratives of $p_x$ with respect to $p_y$ by differentiating the energy relation~(\ref{eq:E-class-trigwarp-dimless}).
  Fixing the energy $E$ and considering $p_x$ to be a function of $p_y$, we obtain
  \begin{equation}
    \frac{\partial E^\pm_\alpha}{\partial p_x} \frac{\partial p_x}{\partial p_y} + \frac{\partial E^\pm_\alpha}{\partial p_y} = 0.
    \label{eq:diff-energy-implicit}
  \end{equation}
  Since we consider $\theta=0$, the second term vanishes at $p_y=0$, and we find that $\partial p_x/\partial p_y$ equals zero at that point. Of course, we could already have concluded this from its antisymmetry in $p_y$.
  Taking the derivative of the above result, and using that the first derivative of $p_x$ vanishes, we obtain
  \begin{equation}
    \frac{\partial^2 p_x}{\partial p_y^2} = - \frac{\partial^2 E^\pm_\alpha/\partial p_y^2}{\partial E^\pm_\alpha/\partial p_x} ,
  \end{equation}
  where all derivatives are to be evaluated at zero $p_y$. Evaluating the energy derivatives using Eq.~(\ref{eq:H-class-trigwarp-dimless}), we find
  \begin{equation}
    \frac{\partial E^\pm_\alpha}{\partial p_x} = \pm \frac{p_x + 2\alpha \mu p_x^2}{|p_x|}, \qquad \frac{\partial^2 E^\pm_\alpha}{\partial p_y^2} = \pm \frac{1- 7 \alpha \mu p_x}{|p_x|} , \qquad 
    \frac{\partial^2 p_x}{\partial p_y^2} = -\frac{1 - 7 \alpha \mu p_x}{p_x + 2 \alpha \mu p_x^2} .
    \label{eq:d2pxdpy2}
  \end{equation}
  We note that the last expression does not contain a $\pm$-sign and is therefore the same for both electrons and holes.
  
  To find the position of the cusp, we need to know the momenta $p_{x,e,\alpha}$ and $p_{x,h,\alpha}$ of the right-moving charge carriers. For electrons, for which me need to take the plus sign in Eq.~(\ref{eq:H-class-trigwarp-dimless}), we set $U_0=0$, require that $p_x>0$ and set $\phi_{\mathbf{p}}=0$. We then obtain
  \begin{equation}
    p_{x,e,K} = \frac{1}{2\mu} \left( 1 - \sqrt{1-4\mu E} \right) , \qquad p_{x,e,K'} = \frac{1}{2\mu} \left( -1 + \sqrt{1+4\mu E} \right) ,
    \label{eq:px-el}
  \end{equation}
  where we have used that for low energies the leading term in the Taylor expansion should give $p_{x,e,\alpha}=E$. 
  For holes, we take the minus sign in Eq.~(\ref{eq:H-class-trigwarp-dimless}) and set $U=U_0>E$, $p_x<0$ and $\phi_{\mathbf{p}}=\pi$. Requiring that $p_{x,h,\alpha}=U_0-E$ in the zeroth order expansion in $\mu$, we have  
  \begin{equation}
    p_{x,h,K} = \frac{1}{2\mu} \left( 1 - \sqrt{1+4\mu(U_0-E)} \right) , \qquad p_{x,h,K'} = \frac{1}{2\mu} \left( -1 + \sqrt{1-4\mu(U_0-E)} \right)
    \label{eq:px-h}
  \end{equation}
  Inserting the results~(\ref{eq:px-el}) and~(\ref{eq:px-h}) into Eq.~(\ref{eq:x-caustic}), we obtain an expression for $x_{\text{cusp},\alpha}$. This expression is rather cumbersome, but can be easily implemented into a computer algebra system. It is used to generate the cusp lines in Fig.~\ref{fig:real-space-wavefunction-maxima}(d)--(f).
  
  To gain more insight into the splitting, we perform a first order Taylor expansion in $\mu$ and obtain
  \begin{equation}
    x_{\text{cusp},\alpha} = L_1 \frac{U_0-E}{E} \left( 1 - 8 \alpha \mu U_0 \right) + \mathcal{O}(\mu^2) .
    \label{eq:x-cusp-trig-expansion}
  \end{equation}
  For small $\mu$, we therefore find the distance between the two cusp points as
  \begin{equation}
    \Delta x_{\text{cusp}} = x_{\text{cusp},K} - x_{\text{cusp},K'} = 16 \mu L_1 U_0 \frac{U_0-E}{E} + \mathcal{O}(\mu^2) .
  \end{equation}
  Going back to units with dimensions, we have, up to second order in $\mu$,
  \begin{equation}
    x_{\text{cusp},\alpha} = L_1 \frac{U_0-E}{E} \left( 1 - \alpha \frac{4 U_0}{3 t} \right) . \quad
  \end{equation}
  and the distance between the two cusp points equals
  \begin{equation}
    \Delta x_{\text{cusp}} = x_{\text{cusp},K} - x_{\text{cusp},K'} = L_1 \frac{8 U_0}{3 t} \frac{U_0-E}{E} .
  \end{equation}
  The implications of these equations have been extensively discussed in the main text.

  \subsection{Position of the maximum}
  
  Within the Dirac approximation, one can construct an analytical solution for scattering by a sharp \emph{n-p} junction~\cite{Cheianov07,Reijnders17}. The wavefunction in the hole region can be expressed as an integral over the transversal momentum $p_y$, which labels the trajectories:
  \begin{equation}
    \Psi(x,y) = \int_{-\infty}^\infty \text{d} p_y \, f(x,y,p_y) e^{i S_{np}(x,y,p_y)/h} . \label{eq:int-generic}
  \end{equation}
  The action $S_{np}$ is given by Eq.~(\ref{eq:action-dimless-app}), with $\mu=0$ for the Dirac case, and the amplitude function $f(x,y,p_y)$ depends on the details of the source~\cite{Reijnders17}. This expression is not only valid for the Green's function, but also for current injection through a lead.
  
  Based on general, partially semiclassical, principles, expression~(\ref{eq:int-generic}) should still be valid when we include trigonal warping. However, in this case, we have $\mu\neq 0$ in the action $S_{np}$ and the amplitude $f(x,y,p_y)$ is likely a more complicated function.
  Within the semiclassical approximation, which is valid for small $h$, we can compute the maximum of the wavefunction~(\ref{eq:int-generic}) near a cusp point using the Pearcey approximation~\cite{Connor81b,Reijnders17,Dobrokhotov14}. Although this approximation typically predicts too large values for the wavefunction, it accurately predicts the position of the maximum, even for relatively large values of $h$~\cite{Reijnders17}. A key observation is that, within this approximation, the position of the maximum does not depend on the function $f(x,y,p_y)$, but only on the action $S_{np}$. When we include trigonal warping, we can therefore obtain a reliable result for the position of the main focus without solving the full problem. However, this approximation is only valid when we consider sufficiently large values of $a_4$. Otherwise, we are dealing with a higher order singularity, and the wavefunction should be approximated using a different special function.
  
  Before applying the Pearcey approximation to our problem, let us briefly review it and introduce the relevant notations.
  The main contribution to the integral~(\ref{eq:int-generic}) is given by the critical points, where $\partial S_{np}/\partial p_y$ vanishes. This gives rise to the classical trajectories, which we discussed in the previous subsection. At the caustic, the second derivative also vanishes, and at the cusp the third derivative is also identically zero:
  \begin{equation}
    \frac{\partial S_{np}}{\partial p_y} = \frac{\partial^2 S_{np}}{\partial p_y^2} = \frac{\partial^3 S_{np}}{\partial p_y^3} = 0 .
    \label{eq:def-cusp-action-derivs}
  \end{equation}  
  We obtain the simplest approximation to the wavefunction~(\ref{eq:int-generic}) near a cusp point by making a Taylor expansion of the action $S_{np}$ around the critical point $(x_{\text{cusp}}, y_{\text{cusp}}, p_{y,0})$ up to the first nonzero term. This gives
  \begin{equation}
    S_{np}(x,y,p_y) = S^{(4)}(x,y,p_y) + \mathcal{O}(\beta^5) = q_0(x,y) + q_1(x,y) \beta + \frac{q_2(x,y)}{2} \beta^2 + \frac{q_3(x,y)}{6} \beta^3 + \frac{q_4(x,y)}{24} \beta^4 + \mathcal{O}(\beta^5)
     , \label{eq:S-exp-Pearcey}
  \end{equation}
  where $\beta=p_y-p_{y,0}$. Because of Eq.~(\ref{eq:def-cusp-action-derivs}), the expansion coefficients satisfy:
  \begin{equation}
    \begin{aligned}
      q_0(x,y) &= a_0 + \langle \mathbf{b}_0 , \mathbf{z} \rangle + \mathcal{O}(z^2), &
      q_1(x,y) &= \langle \mathbf{b}_1 , \mathbf{z} \rangle + \mathcal{O}(z^2), &
      q_2(x,y) &= \langle \mathbf{b}_2 , \mathbf{z} \rangle + \mathcal{O}(z^2), \\
      q_3(x,y) &= \mathcal{O}(z) &
      q_4(x,y) &= a_4 + \mathcal{O}(z) . 
      \label{eq:q-exp-Pearcey}
    \end{aligned}
  \end{equation}
  where $\mathbf{z}=\mathbf{x}-\mathbf{x}_0 = (x,y)-(x_{\text{cusp}},y_{\text{cusp}})$.
  
  The leading order approximation to the wavefunction can then be expressed~\cite{Connor81b,Reijnders17,Dobrokhotov14} in terms of the Pearcey function $\text{P}^\pm(x)$, which is defined by the integral
  \begin{equation}
    \text{P}^{\pm}(u,v) = \int_{-\infty}^\infty \exp\left( \pm i t^4 + i u t^2 + i v t \right) \, \text{d} t , \label{eq:def-Pearcey}
  \end{equation}
  where the superscript plus or minus corresponds to the sign in front of the $t^4$ term. 
  This function possesses two important symmetries~\cite{Connor81a,Connor82}, which simplify computations:
  \begin{equation}
    \text{P}^\pm(u,-v)=\text{P}^\pm(u,v), \quad \text{P}^-(u,v) = [\text{P}^+(-u,-v)]^* . \label{eq:symm-Pearcey}
  \end{equation}
  These can be verified directly using definition~(\ref{eq:def-Pearcey}). Furthermore, 
  \begin{equation}
    \text{P}_v^\pm(u,-v) = -\text{P}_v^\pm(u,v), \quad \text{P}_u^\pm(u,-v) = \text{P}_u^\pm(u,v) ,
  \end{equation}
  where $\text{P}_v^\pm(u,v)$ is the partial derivative with respect to the second argument, and $\text{P}_u^\pm(u,v)$ the partial derivative with respect to the first argument.
  Numerical values of the Pearcey function can be computed rather efficiently by deforming the integration contour~\cite{Connor82} or by numerically solving a differential equation~\cite{Connor81a}.
  
  Within the Pearcey approximation, one then obtains the following approximation to the integral~(\ref{eq:int-generic})~\cite{Connor81b,Reijnders17,Dobrokhotov14}:
  \begin{equation}
    \Psi(x,y)= f(x_{\text{cusp}},y_{\text{cusp}},p_{y,0}) \sqrt[4]{\frac{24 h}{|a_4|}} \exp\left[\frac{i}{h}\left( a_0 + \langle \mathbf{b}_0 , \mathbf{z} \rangle \right)\right] \\ 
    \text{P}^{\pm} \left[ \sqrt{\frac{6}{h |a_4|}} \langle \mathbf{b}_2 , \mathbf{z} \rangle , \sqrt[4]{\frac{24}{h^3 |a_4|}} \langle \mathbf{b}_1 , \mathbf{z} \rangle \right] + \mathcal{O}(h^{1/2}) , \label{eq:Pearcey-caustic}
  \end{equation}
  where the coefficients are defined by Eq.~(\ref{eq:q-exp-Pearcey}). Taking the norm of this approximation, we indeed see that the position of its maximum does not depend on the function $f$, but only on the expansion coefficients of the action, i.e. on purely classical quantities.
  We see from Eq.~(\ref{eq:symm-Pearcey}) that the Pearcey function is symmetric, and a numerical study shows that its maximum lies on the $x$-axis.
  One can show explicitly that~\cite{Connor81a}
  \begin{equation}
    P^+(x,0) = \frac{\pi |x|^{1/2} \exp(-i x^2/8)}{4 \sin(\pi/4)} \left[ \exp(i \pi/8) J_{-1/4}(x^2/8) - \text{sign}(x) \exp(-i\pi/8) J_{1/4}(x^2/8) \right] ,
  \end{equation}
  where $J_\alpha(x)$ is the Bessel function.
  The maximum of $| P^+(x,0) |$ can be determined numerically and lies at $x_{0,+}=-2.19863$. By the symmetry relations~(\ref{eq:symm-Pearcey}), we therefore find that the maximum of $|P^\pm(x,0)|$ lies at $x_{0,\pm}=\mp 2.19863$.
  Therefore, when $a_4$ is sufficiently large, we can find a good approximation for the position of the maximum by solving the set of equations
  \begin{equation}
    \sqrt{\frac{6}{h |a_4|}} \langle \mathbf{b}_2 , \mathbf{x}_{\text{max}}-\mathbf{x}_{\text{cusp}} \rangle = x_{0,\pm} , \qquad
    \sqrt[4]{\frac{24}{h^3 |a_4|}} \langle \mathbf{b}_1 , \mathbf{x}_{\text{max}}-\mathbf{x}_{\text{cusp}} \rangle = 0  .
    \label{eq:max-Pearcey-eqs}
  \end{equation}

  We now return to the case $\theta=0$, for which we explicitly computed the position of the caustic in the previous section. The coefficient $a_{4,\alpha}$ is given by
  \begin{equation}
    a_{4,\alpha} = L_1 \frac{\partial^4 p_{x,e,\alpha}}{\partial p_y^4} + x_{\text{cusp},\alpha} \frac{\partial^4 p_{x,h,\alpha}}{\partial p_y^4} ,
    \label{eq:a4-abstract}
  \end{equation}
  where all quantities are to be evaluated at $p_y=0$.
  We can obtain an expression for the fourth derivative of $p_x$ in the same way as we obtained an expression for the second derivative in Eqs.~(\ref{eq:diff-energy-implicit})--(\ref{eq:d2pxdpy2}), using the fact that the odd order derivatives of $E_\alpha^\pm$ with respect to $p_y$ vanish. We find that
  \begin{equation}
    \frac{\partial^4 p_x}{\partial p_y^4} = - 3 \frac{1 + 3 \alpha \mu p_x + 158 \alpha \mu^3 p_x^3}{p_x^3 (1 + 2 \alpha \mu p_x)^3}
  \end{equation}
  Inserting this into Eq.~(\ref{eq:a4-abstract}), we obtain a complex expression for $a_{4,\alpha}$. Performing a Taylor expansion in $\mu$, we find
  \begin{equation}
    a_{4,\alpha} = - 3 L_1 \frac{U_0}{E^3} \frac{U_0 - 2 E}{(U_0-E)^2} - 24 \alpha \mu L_1 \frac{U_0}{E(U_0-E)^2} + \mathcal{O}(\mu^2) 
    \stackrel{U_0=2E}{=\mathrel{\mkern-3mu}=\mathrel{\mkern-3mu}=\mathrel{\mkern-3mu}=} -48 \alpha \mu \frac{L_1}{E^2} + \mathcal{O}(\mu^2) 
    \label{eq:a4-expansion}
  \end{equation}
  At $U_0=2E$, the term that is zeroth order in $\mu$ vanishes. We then have $a_{4,K} > 0$, which means that the maximum lies left of the cusp point for the $K$-valleys, and $a_{4,K'} < 0$, i.e. the maximum lies right of the cusp point for the $K'$-valley. In both cases, the maximum lies closer to the point $x=L_1$ than the cusp point.

  The coefficients $\mathbf{b}_{n,\alpha}$ in Eq.~(\ref{eq:q-exp-Pearcey}) can be computed as
  \begin{equation}
    b_{n,\alpha,x} = \left. \frac{\partial}{\partial x} \frac{\partial^n S_{np}}{\partial p_y^n} \right|_{\mathbf{x}=\mathbf{x}_\text{cusp}} , \qquad
    b_{n,\alpha,y} = \left. \frac{\partial}{\partial y} \frac{\partial^n S_{np}}{\partial p_y^n} \right|_{\mathbf{x}=\mathbf{x}_\text{cusp}} .
  \end{equation}
  Specializing to $\theta=0$, we obtain
  \begin{equation}
    b_{1,\alpha,x} = 0 , \qquad b_{1,\alpha,y} = 1 , \qquad 
    b_{2,\alpha,x} = \frac{\partial^2 p_{x,h,\alpha}}{\partial p_y^2} , \qquad 
    b_{2,\alpha,y} = 0 ,
    \label{eq:b-ZZ}
  \end{equation}
  where we have used that $\partial p_x/\partial p_y = 0$.
  We can make an expansion of $b_{2,\alpha,x}$ and find
  \begin{equation}
    b_{2,\alpha,x} =  \frac{1}{U_0 - E} + 8 \alpha \mu + \mathcal{O}(\mu^2)
  \end{equation}
  Combining Eqs.~(\ref{eq:max-Pearcey-eqs}) and~(\ref{eq:b-ZZ}), we observe that the maximum is on the $x$-axis, and that its position can be found by solving 
  \begin{equation}
    \sqrt{\frac{6}{h |a_{4,\alpha}|}} b_{2,\alpha,x} (x_{\text{max},\alpha} - x_{\text{cusp},\alpha} ) = x_{0,\pm}
    \label{eq:x-max-Pearcey-defining}
  \end{equation}
  Hence, the position of the maximum equals
  \begin{equation}
    x_{\text{max},\alpha} = x_{\text{cusp},\alpha} + \frac{x_{0,\pm}}{b_{2,\alpha,x}} \sqrt{\frac{h |a_{4,\alpha}|}{6}}
    \label{eq:x-max-Pearcey}
  \end{equation}
  We emphasize that we cannot use this approximation for low energies (small $\mu$) at $U_0=2E$, since $a_{4,\alpha}$ is small in that case, which implies that we are considering a section of a higher order caustic. In fact, we are too close to the Dirac case, where an ideal focus is observed at $U_0=2E$.

  Finally, let us briefly return to units with dimensions and investigate the dependence of this expression on $L_1$ at constant $E$ and $U_0$.
  When restoring dimensions, the dimensionless variables $\tilde{x}_{\text{max},\alpha}$ and $\tilde{x}_{\text{cusp},\alpha}$, which are used in Eq.~(\ref{eq:x-max-Pearcey}), are replaced by $x_{\text{max},\alpha}/L_1$ and $x_{\text{cusp},\alpha}/L_1$.
  Furthermore, we note that $x_{\text{cusp},\alpha}$ scales linearly with length. The coefficient $a_{4,\alpha}$ is therefore independent of the length scale of the system, as is $b_{2,\alpha,x}$. Hence, the only part of the second term that scales with length is the semiclassical parameter $h$, which is proportional to $1/L_1$.
  Therefore, we conclude that the dependence of $x_{\text{max},\alpha}$ on $L_1$ is given by 
  \begin{equation}
    x_{\text{max},\alpha}(L_1) = c_1 L_1 + c_2 \sqrt{L_1} ,
  \end{equation}
  where $c_1$ and $c_2$ are constants that do not depend on $L_1$. This behavior is indeed observed in Fig.~\ref{fig:real-space-wavefunction-maxima}(e).

  \subsection{Higher order singularities}
  
\begin{figure*}[!ptbh]
  \includegraphics{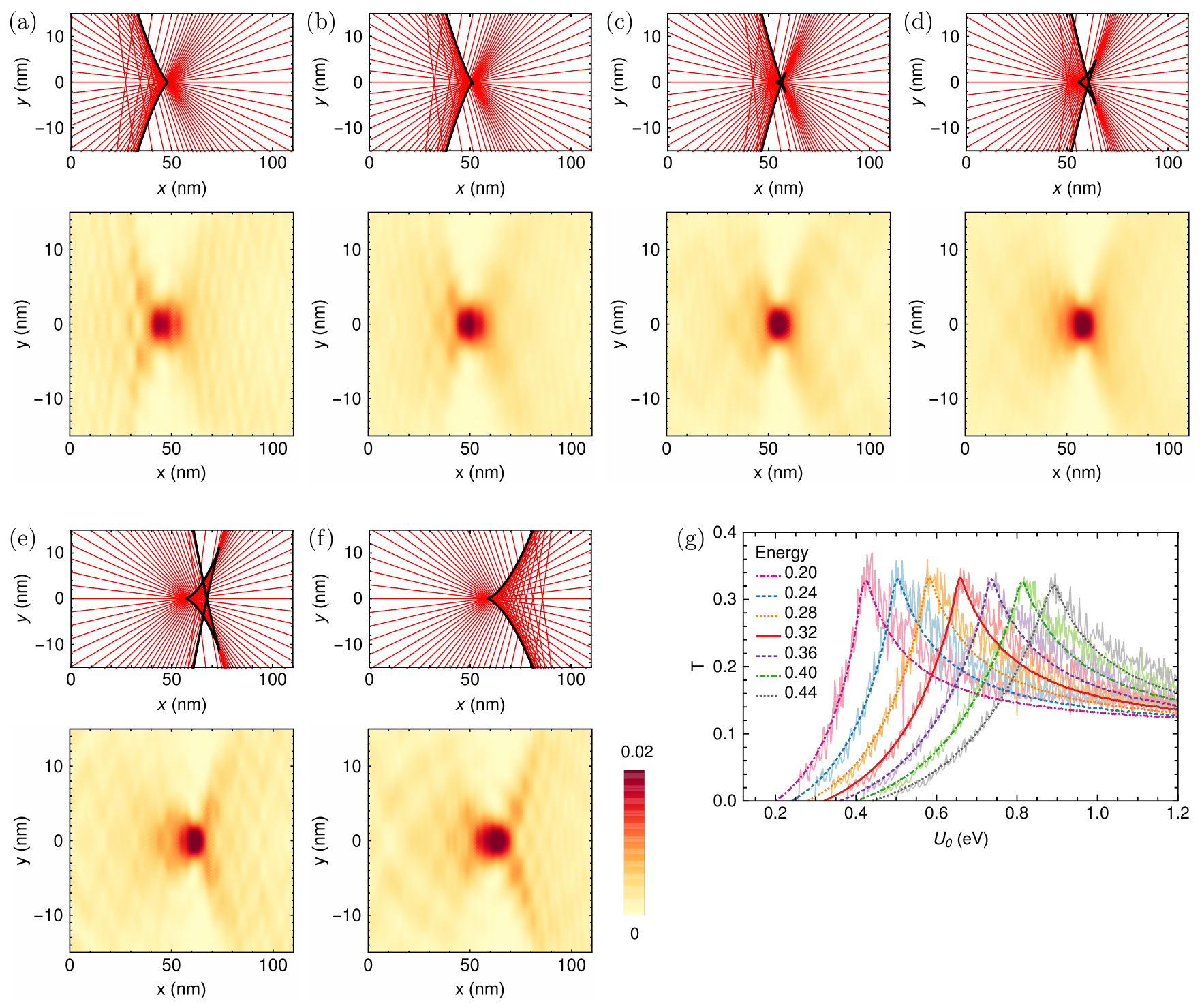}
  \caption{(a)--(f) Evolution of the classical paths and the wavefunction computed with Kwant as we pass through the section of the butterfly caustic in the $K'$-valley. We consider a sample with $W_i=7.5$~nm, $L_1=100$~nm and $E=0.6$~eV. For this energy, the coefficient $a_{4,K'}$ vanishes at $U_0=1.069$ eV. The potentials $U_0$ are equal to (a) 1.07~eV, (b) 1.10~eV, (c) 1.14~eV, (d) 1.16~eV, (e) 1.18~eV, (f) 1.20~eV. Increasing the potential, we see that the side at which the interference pattern is visible changes from left to right and that we pass through a sharper focus.
  (g) Evolution of the transmission for the $K$-valley as a function of potential strength as we pass through the butterfly caustic by changing the electron energy. We consider a sample with $W_i=50$~nm, $L_1=100$~nm and $L_2=140$~nm. The darker (smooth) lines correspond to the results of the billiard model, and the lighter (fuzzy) lines correspond to the result of the Kwant simulations. Computing the potential $U_{0,\text{hs},K}$ for which $a_{4,K}$ vanishes for a given energy and the position~$x_{\text{hs},K}$ of the butterfly singularity for this potential, we find that $x_{\text{hs},K}=140$~nm and $U_{0,\text{hs},K}=0.5$~eV for $E=0.24$~eV. This means that the classical trajectories for that energy and potential look approximately like those shown in figure~(a), although they are mirrored in a line parallel to the $y$-axis. When we increase the energy, the set of classical trajectories at the point of maximal transmission looks like a mirrored version of figures (c), (d) and (e). In the billiard model, the maximal transmission is slightly higher at $E=0.32$~eV, but this is not observed in the Kwant data.}
  \label{fig:realspacebutterfly}
  \end{figure*}

  For $\theta=0$, let us also consider the points where $a_{4,\alpha}$ vanishes and at which we pass through a higher order singularity. In the simplest approximation, we can analyze these points by looking at the first order Taylor expansion in $\mu$, Eq.~(\ref{eq:a4-expansion}). For the $K$-valley, the term that is linear in $\mu$ is always positive, whereas it is always negative for the $K'$-valley. The first term does not depend on the valley index, but is negative for $U_0>2E$, and positive for $U_0<2E$. Therefore, the coefficient $a_{4,\alpha}$ vanishes at $U_0>2E$ for the $K$-valley and at $U_0<2E$ for the $K'$-valley. Neglecting terms of $\mathcal{O}(\mu^2)$ in the expansion of $a_{4,\alpha}$ and solving for the potential $U_{0,\text{hs},\alpha}$ where it vanishes, we obtain
  \begin{equation}
    U_{0,\text{hs},\alpha} = 2 E - 8 \alpha \mu E^2 .
    \label{eq:U0-hs}
  \end{equation}
  When we consider the Dirac Hamiltonian, i.e. $\mu=0$, this reduces to $U_{0,\text{hs},\alpha}=2E$, as we already extensively discussed.
  Let us also compute the sixth order derivative of the action at the cusp, to see whether it is nonzero at the points where $a_{4,\alpha}$ vanishes. We have
  \begin{equation}
    a_{6,\alpha} =  \frac{\partial^6 S_{np}}{\partial p_y^6} = L_1 \frac{\partial^6 p_{x,e,\alpha}}{\partial p_y^6} + x_{\text{cusp},\alpha} \frac{\partial^6 p_{x,h,\alpha}}{\partial p_y^6} ,
    \label{eq:a6}
  \end{equation}
  where the derivatives are to be evaluated at zero $p_y$. Using the same procedure as before to compute the sixth derivative of $p_x$ with respect to $p_y$ at $p_y=0$, we obtain 
  \begin{equation}
    \frac{\partial^6 p_x}{\partial p_y^6} = 45 \frac{-1 - 5 \alpha \mu p_x  + 108 \mu^2 p_x^2 + 
    796 \alpha \mu^3 p_x^3 + 1808 \mu^4 p_x^4 + 
    4260 \alpha \mu^5 p_x^5}{p_x^5 (1 + 2 \alpha \mu p_x)^5}
  \end{equation}
  Inserting the momenta~(\ref{eq:px-el}) and~(\ref{eq:px-h}) into this result, we obtain an expression for $a_{6,\alpha}$. We can then specialize to the point $U_{0,\text{hs},\alpha}$, where $a_{4,\alpha}$ vanishes. Inserting the result~(\ref{eq:U0-hs}) and expanding the result up to first order in $\mu$, we obtain
  \begin{equation}
    a_{6,\text{hs},\alpha} = 720 \alpha \mu \frac{L_1}{E^4} .
  \end{equation}
  We see that this coefficient is indeed nonzero when $\mu$ is nonzero, indicating that the butterfly singularity is the highest type of singularity that can be expected in the system when trigonal warping is included. This is confirmed by a numerical analysis, where we numerically solve for $U_{0,\text{hs},\alpha}$ without neglecting higher order terms in $\mu$ and insert the result into $a_{6,\alpha}$.
  For the $K'$-valley, one then has $a_{6,\text{hs},\alpha}>0$ and for the $K$-valley, one has $a_{6,\text{hs},\alpha}<0$. The caustics that are observed in the set of classical trajectories, see Figs.~\ref{fig:setup-trajectories}(e) and~\ref{fig:realspacebutterfly}, agree with this conclusion~\cite{Poston78}.

  Determining the position of the maximum for this higher order singularity requires a special function involving $\exp(iu^6)$ in the integrand. We will not go into this here, since the generic singularity in our system is the cusp. Instead, we determine the position of the butterfly singularity. Entering $U_{0,\text{hs},\alpha}$ into Eq.~(\ref{eq:x-caustic}) and expanding the result up to first order in $\mu$, we obtain
  \begin{equation}
    x_{\text{hs},\alpha} = L_1 ( 1 - 24 \alpha \mu E ) ,
  \end{equation}
  which reduces to the Dirac result $x_{\text{hs},\alpha}=L_1$ when we set $\mu =0$.

  \subsection{Position of the maximum of the transmission peak when varying $U_0$}
  
  Let us now consider the case $\theta=0$ and have a look at the transmission as a function of potential strength. Fixing $L_1$ and $L_2$ and varying the potential $U_0$, one finds a value $U_{0,\text{max},\alpha}$ at which the transmission is maximal. We expect this maximum to be reached when the position $x_{\text{max},\alpha}$ of the maximum equals $L_2$. Therefore, we can find an approximation to $U_{0,\text{max},\alpha}$ by replacing $x_{\text{max},\alpha}$ by $L_2$ in Eq.~(\ref{eq:x-max-Pearcey-defining}) and by subsequently solving for $U_0$.
  Since it is not easy to solve this equation analytically, we use numerical methods to find a solution.
  
  Alternatively, we can consider the potential $U_{0,\text{cusp},\alpha}$ for which $x_{\text{cusp},\alpha}$ equals $L_2$. This will give us an estimate of the potential at which the maximum is reached. This estimate will generally be too low (high) for the valley $K$ ($K'$), since the maximum in real space lies to the left (right) of the cusp point. We remark that when $L_1=L_2$ and we are considering low energies, this is the only method with which we can say something about the position of the maximum, as Eq.~(\ref{eq:x-max-Pearcey-defining}) is not valid for this case due to the smallness of $a_{4,\alpha}$. Looking at Eq.~(\ref{eq:x-caustic}), we see that the only dependence on the length scale of the system is through $L_2/L_1$. Since these are dimensionless quantities, and their ratio is independent of the length scale of the system, our first conclusion is that $U_{0,\text{cusp},\alpha}$ does not depend on this length scale.
  Let us now compute $U_{0,\text{cusp},\alpha}$ for low energies, which correspond to small values of $\mu$. We can then use the expansion~(\ref{eq:x-cusp-trig-expansion}) for $x_{\text{cusp},\alpha}$ and solve the equation $x_{\text{cusp},\alpha}=L_2$ for $U_0$. Expanding the final result for $U_{0,\text{cusp},\alpha}$ in $\mu$, we obtain
  \begin{equation}
    U_{0,\text{cusp},\alpha} = E \left( 1 + \frac{L_2}{L_1} \right) \left( 1 + 8 \alpha \mu E \frac{L_2}{L_1} \right) + \mathcal{O}(\mu^2)
    \stackrel{L_1=L_2}{=\mathrel{\mkern-3mu}=\mathrel{\mkern-3mu}=\mathrel{\mkern-3mu}=} 2 E + 16 \alpha \mu E^2 + \mathcal{O}(\mu^2) .
    \label{eq:U0-cusp}
  \end{equation}
  For the Dirac case, where $\mu = 0$, we obtain $U_{0,\text{cusp},\alpha} = E ( 1 + L_2/L_1 )$, which is the solution of the equation $L_1 (U_0-E)/E = L_2$ that one easily establishes for the Dirac equation. From Eq.~(\ref{eq:U0-cusp}), we then immediately see that the maximum for the $K$-valley occurs at a lower potential than for the Dirac equation, whereas the maximum for the $K'$-valley occurs at a higher potential than for the Dirac equation.

  \subsection{Height of the transmission peak when varying $U_0$}
  
  We also investigate the maximal height of the transmission peak that is obtained by varying $U_0$, in order to see whether it differs for the valleys $K$ and $K'$. Although the Pearcey approximation, Eq.~(\ref{eq:Pearcey-caustic}), generally predicts too high values~\cite{Reijnders17} for the absolute value of the wavefunction when we consider small $h$, we may be able to obtain some qualitative information about this difference.
  
  Looking at Eq.~(\ref{eq:Pearcey-caustic}), we see that the peak height is proportional to the inverse of $|a_{4,\alpha}|$ and to the amplitude function~$f(x_{\text{cusp},\alpha},y_{\text{cusp},\alpha},p_{y,0})$, which should be evaluated at $U_{0,\text{max},\alpha}$.
  However, we do not expect that the latter function will significantly contribute to a difference in peak height, considering its functional form for the Dirac equation, see Ref.~\cite{Reijnders17}.
  For the Dirac equation, the function $f$ essentially consists of two parts. The first part comes from the electron source and depends on the angle of incidence and the sublattice polarization. The second part is the transmission coefficient through the barrier, which is to be evaluated at $p_{y,0}$. Because of the symmetry of the system, $p_{y,0}=0$ for both valleys. In the Dirac picture, this corresponds to normal incidence, i.e. the incidence angle $\phi=0$, and hence all phase factors that depend on the angle equal unity. Furthermore, at normal incidence we have Klein tunneling, i.e. the transmission is unity regardless of the height of the potential.
  Although trigonal warping certainly introduces corrections in the amplitude function, we do not expect that these will greatly influence either of the two factors at the cusp point. Therefore, we believe that the amplitude function $f$ at the cusp point depends only weakly on both the potential $U_0$ and the valley index $\alpha$. We thus believe that the factor that has the largest influence on a possible difference in peak height between the two valleys is the coefficient $|a_{4,\alpha}|$, on which we therefore focus our attention.
  
  As with the position of the maximum, one generally needs to resort to numerical methods to obtain the coefficient $|a_{4,\alpha}|$ at $U_{0,\text{max},\alpha}$. However, to gain qualitative understanding of the effect, and to be able to obtain an analytical expression, we can also evaluate the coefficient $a_{4,\alpha}$ at $U_{0,\text{cusp}\alpha}$, given by Eq.~(\ref{eq:U0-cusp}). Expanding the result up to first order in $\mu$, we obtain
  \begin{equation}
    a_{4,\alpha} = 3 \frac{L_1}{E^3} \left(\frac{L_1^2}{L_2^2} - 1 \right) - 72 \alpha \mu \frac{L_1}{E^2} \frac{L_1}{L_2} \left( 1 + \frac{L_1}{L_2} \right) + \mathcal{O}(\mu^2) .
  \end{equation}
  For $L_1=L_2$, we see that the first term vanishes. The absolute value of the second term is equal for both values of $\alpha$, implying that, for small values of $\mu$ and hence for low energies, the transmission peaks for both valleys will be equal in height when we consider $L_1=L_2$.
  
  If we want to see whether this changes when we go to higher energies, we need to consider the second term in the expansion of $a_{4,\alpha}$. Unfortunately, when we expand $x_{\text{cusp},\alpha}$ to second order in $\mu$, the equation $x_{\text{cusp},\alpha} = L_2$ becomes third order in $U_0$, which makes the expression for $U_{0,\text{cusp},\alpha}$ much more complicated. This also makes the second order expansion of $a_{4,\alpha}$ in $\mu$ at $U_{0,\text{cusp},\alpha}$ very cumbersome. Nonetheless, we can say something about it by taking another look at our previous equations. The key observation is that in all our equations $\mu$ appears together with $\alpha$ in the combination $\alpha \mu$. This holds for the Hamiltonian~(\ref{eq:H-class-trigwarp-dimless}) and for the momenta~(\ref{eq:px-el}) and~(\ref{eq:px-h}) and therefore also for all derived quantities, such as $x_{\text{cusp},\alpha}$ and $a_{4,\alpha}$. This implies that we are actually dealing with an expansion in $\alpha \mu$ rather than in just $\mu$, which means that the second term in the expansion is proportional to $\alpha^2 \mu^2$. Since $\alpha^2 = 1$, this term breaks the symmetry between the two valleys that is present in the linear term. Therefore, when we include the quadratic term in the expansion, $|a_{4,\alpha}|$ is different for the two values of $\alpha$.
  
  Unfortunately, this analysis does not show whether the sign in front of the second term is positive or negative. A numerical study of the expansion shows that it is generally positive, i.e. that $|a_{4,K}|$ is larger than $|a_{4,K'}|$. We find the same effect when we numerically evaluate $a_{4,\alpha}$ at $U_{0,\text{max},\alpha}$ for typical energies and length scales. This predicts that, for high energies, the height of the transmission peak for the $K$-valley is lower than the height for the $K'$-valley, which is indeed seen in our numerical experiments.
  
  We remark that the situation changes when we consider incident electrons with a uniform angular distribution, rather than a distribution that is uniform in momenta. For a uniform angular distribution, we need to integrate over the incidence angle $\phi$ rather than over the momentum $p_y$. Changing variables, we obtain an additional factor $\partial \phi/\partial p_y$ in the amplitude. It is clear that this does not change the classical trajectories, and that the position and width of the maximum will be the same for both distributions. However, the height of the transmission peak is affected by this factor.
  The angle $\phi$ is determined by the two components of the group velocity of the electrons, i.e.
  \begin{equation}
    \phi(p_y) = \arctan \left( \frac{\partial H_\alpha/\partial p_y}{\partial H_\alpha/\partial p_x} \right) .
    \label{eq:angle-py-trigwarp}
  \end{equation}
  In the Pearcey approximation, we need to evaluate $\partial \phi/\partial p_y$ at $p_{y,0}=0$. After some calculus, using the fact that $\partial H_\alpha/\partial p_y$ vanishes at $p_y=0$, we find
  \begin{equation}
    \frac{\partial \phi}{\partial p_y} \stackrel{p_y=0}{=\mathrel{\mkern-3mu}=\mathrel{\mkern-3mu}=\mathrel{\mkern-3mu}=}
    \frac{\partial^2 H_\alpha/\partial p_y^2}{\partial H_\alpha/\partial p_x} = \frac{1}{E} - 8 \alpha \mu + \mathcal{O}(\mu^2) .
  \end{equation}
  We therefore see that this factor is larger for the $K$-valley than for the $K'$-valley, which could counter the effect of the coefficient $a_4$. Since the values given by the Pearcey approximation are generally too large for small values of $h$, these equations cannot tell us what the combined effect of these two factors will be. Our billiard simulations show that for a uniform angular distribution both peaks are approximately equal in height.

  \subsection{Scaling of the width of the transmission peak}
  
  We have already stated that the maximal transmission will be obtained for the potential $U_{0,\text{max},\alpha}$ for which $x_{\text{max},\alpha}=L_2$. Changing the potential, we obtain a peak in the transmission, which has a full width at half maximum (FWHM) $\Delta U_{0,\alpha}$. Let us now consider the special case $L_2=L_1=L$ and see how this width $\Delta U_{0,\alpha}$ changes when we vary $L$.
  To this end, we first need to find the potential $U_{0,\text{left},\alpha}$, where the transmission equals a certain amount, for instance half, of its maximal value. Let us therefore denote by $x_{1,\pm}$ ($x_{2,\pm}$) the point on the $x$-axis to the left (right) of $x_{0,\pm}$ where the absolute value of the Pearcey function $P^\pm(x,0)$ equals $1/\sqrt{2}$ of its maximal value. By symmetry, these points are related by $x_{1,\pm}=-x_{2,\mp}$. Before, we obtained an estimate for $x_{\text{max},\alpha}$ by solving Eq.~(\ref{eq:x-max-Pearcey-defining}). To obtain $U_{0,\text{left},\alpha}$, we replace $x_{0,\pm}$ in Eq.~(\ref{eq:x-max-Pearcey-defining}) by $x_{2,\pm}$ and change $x_{\text{max},\alpha}$ to $L$. Then $U_{0,\text{left},\alpha}$ is given by the potential $U_0$ that solves this equation for a given energy. Similarly, replacing $x_{0,\pm}$ by $x_{1,\pm}$, we find $U_{0,\text{right},\alpha}$.
  
  By the above procedure, we obtain two equations, namely
  \begin{equation}
    L = x_{\text{cusp},\alpha}(E,U_{0,\text{left},\alpha}) + \sqrt{h} g(E,U_{0,\text{left},\alpha}) x_{2,\pm} \; , \qquad
    L = x_{\text{cusp},\alpha}(E,U_{0,\text{right},\alpha}) + \sqrt{h} g(E,U_{0,\text{right},\alpha}) x_{1,\pm} \; ,
    \label{eq:width-U0}
  \end{equation}
  where
  \begin{equation}
    g(E,U_0) = \frac{1} {b_{2,\alpha,x}} \sqrt{\frac{|a_{4,\alpha}|}{6}} .
  \end{equation}
  We remark that, since we are using dimensionless units, $L=1$ and $x_{\text{cusp},\alpha}$, given by Eq.~(\ref{eq:x-caustic}), does not depend on the length scale of the system. Since both $U_{0,\text{left},\alpha}$ and $U_{0,\text{right},\alpha}$ are close to $U_{0,\text{cusp},\alpha}$, we can perform a first order Taylor expansion of $x_{\text{cusp},\alpha}(E,U_0)$ in $U_0$ around $U_{0,\text{cusp},\alpha}$. When we also expand $g(E,U_0)$ to zeroth order around $U_{0,\text{cusp},\alpha}$, we obtain the set of equations
  \begin{equation}
    \begin{aligned}
      L &= x_{\text{cusp},\alpha}(E,U_{0,\text{cusp},\alpha}) + \left.\frac{\partial x_{\text{cusp},\alpha}}{\partial U_0}\right|_{U_{0,\text{cusp},\alpha}} (U_{0,\text{left},\alpha} - U_{0,\text{cusp},\alpha}) + \sqrt{h} g(E,U_{0,\text{cusp},\alpha}) x_{2,\pm} , \\
      L &= x_{\text{cusp},\alpha}(E,U_{0,\text{cusp},\alpha}) + \left.\frac{\partial x_{\text{cusp},\alpha}}{\partial U_0}\right|_{U_{0,\text{cusp},\alpha}} (U_{0,\text{right},\alpha} - U_{0,\text{cusp},\alpha}) + \sqrt{h} g(E,U_{0,\text{cusp},\alpha}) x_{1,\pm} .
    \end{aligned}
  \end{equation}
  We note that $U_{0,\text{cusp},\alpha}$ solves the equation $x_{\text{cusp},\alpha}(E,U_{0})=L$, whence the first term on the right-hand side cancels the term on the left-hand side in both equations.
  Subtracting the two equations above, and performing some elementary operations, we find that
  \begin{equation}
    \Delta U_{0,\alpha} = U_{0,\text{right},\alpha} - U_{0,\text{left},\alpha} = \frac{g(E,U_{0,\text{cusp},\alpha})}{\partial x_{\text{cusp},\alpha}/\partial U_0} ( x_{2,\pm} -  x_{1,\pm} ) \sqrt{h} 
    \label{eq:U0-width-l-dep}
  \end{equation}
  Since the predictions given by the Pearcey approximation are usually too large, we do not expect the estimate~(\ref{eq:U0-width-l-dep}) to be very precise. However, it does give us information on the scaling behavior of $\Delta U_{0,\alpha}$ with $L$. Since $U_{0,\text{cusp},\alpha}$ does not depend on the length scale of the system, as we established above, the only dependence on length is in the semiclassical parameter $h$. Therefore, we conclude that, as we vary the potential $U_0$, the width of the transmission peak scales as $L^{-1/2}$.
  
  One may object that the scaling behavior expressed in Eq.~(\ref{eq:U0-width-l-dep}) was obtained using a first order Taylor expansion of $x_{\text{cusp},\alpha}$ and a zeroth order expansion of $g(E,U_0)$.
  When we also include the first-order term in the Taylor expansion of $g(E,U_0)$ around $U_{0,\text{cusp},\alpha}$, the expression for $\Delta U_{0,\alpha}$ becomes more complicated. However, when we expand the result, we still find that the leading-order term is proportional to $L^{-1/2}$, although we also obtain corrections of order $L^{-1}$ and higher. This statement remains true when we add the second-order terms in both Taylor expansions.
  Therefore, the leading-order term of the width of the transmission peak scales as $L^{-1/2}$.
  
  Finally, let us briefly say something about the difference in peak width between the $K$ and $K'$ valleys. It is hard to perform an analytical study of the Eqs.~(\ref{eq:width-U0}) to find an analytical expression for the width of the peak. However, we can study their solutions numerically for both valleys. Because of the limitations of the Pearcey approximation, the width that is predicted by such a study is probably not very accurate. 
  However, a numerical analysis does predict that the peak width will generally be larger for the $K'$-valley than for the $K$-valley, in agreement with our numerical experiments.

  \clearpage

  \section{Details of the numerical simulations and data processing}
  
  This section contains additional information on the numerical implementation and the way the data were processed.

  \subsection{Setup and data processing for the simulations using the Kwant code}
  
  For the numerical simulations using the Kwant code~\cite{Groth14}, we create a sample using the setup shown in Fig.~\ref{fig:setup-trajectories}(a). It consists of an injector lead of width $W_i$ through which the particles enter and a collector lead of width $W_c$ through which they exit. Particles can also exit through the two drain leads on the side, which diminish scattering by the sides of the sample and thereby prevent infinite internal reflections. The injector lead is modelled as a semi-infinite ribbon, with translation vector $m \mathbf{v}_1 + n \mathbf{v}_2$, where $\mathbf{v}_1$ and $\mathbf{v}_2$ are the lattice vectors defined in Eq.~(\ref{eq:lattice-vecs}). In our standard setup, $m=1$ and $n=0$, which means that the leads have zigzag edges along the translationally invariant direction. When $n\neq 0$, the translation vector makes an angle $\theta$ with the vector $\mathbf{v}_1$. This angle corresponds to the angle $\theta$ of the passive rotation that we considered in the detailed derivations. When $\theta=0$, which means that we are in the standard setup, we can easily separate the modes in the $K$ and the $K'$-valley by their momentum~\cite{Brey06}. When we consider rotated samples, we only consider those angles $\theta$ for which we can clearly separate the modes in different valleys. For $\theta=0$, we have checked that the lead wavefunctions obtained from the Kwant code are in agreement with the solutions of the continuum Dirac model~\cite{Brey06}.
  
  For the computations of the wavefunction in real space, $W_c$ equals $W$, the width of the sample. We choose the length $L_2$ from the junction interface to the collector lead sufficiently long, and have checked that it does not significantly influence the outcomes. The sample width $W$ was typically chosen as $2 L_1$, with $L_1$ the distance from the injector lead to the junction interface. We consider this width realistic and at the same time sufficiently large to not substantially influence the outcome. From a computation of the $S$-matrix, we conclude that intervalley scattering is present, but is not the dominant scattering process. The amount of intervalley scattering, measured as a fraction of the total transmission into the collector lead, is larger for lead modes with higher transversal momentum, most likely due to interaction with the drain leads at the edges. It ranges from approximately 1\% for the lowest lead mode, which is almost completely transmitted into the collector lead, up to 60\% for higher lead modes, which are largely transmitted into the drain leads and therefore have a much smaller effect on the outcome. We come back to the size of intervalley scattering and provide a more quantitative picture later on, when we discuss scattering as a function of potential strength.
  When drain leads are absent, the position of the focus is somewhat changed, but valley splitting is still observed. We ascribe the change of position to additional internal reflections in the electron region, which make their way into the hole region and cause additional interference.
  
  In order to be able to compare our computed wavefunction with semiclassical results, we do not plot the square of the absolute value of the wavefunction on each site, but instead average it over the two sublattices to obtain the envelope function. We do this, for each mode separately, by averaging the wavefunction over a site and its three neighboring sites. Numerically, we replace $|\Psi_{i,\alpha,n}|^2$, with $\Psi_{i,\alpha,n}$ the wavefunction on site $i$ resulting from lead mode $n$ in valley $\alpha$, by $|\Psi_{i,\text{av},\alpha,n}|^2 = \tfrac{1}{2} |\Psi_{i,\alpha,n}|^2 + \tfrac{1}{6} \sum_{\langle i j \rangle} |\Psi_{j,\alpha,n}|^2$. 
  
\begin{figure*}[!ptbh]
  \includegraphics{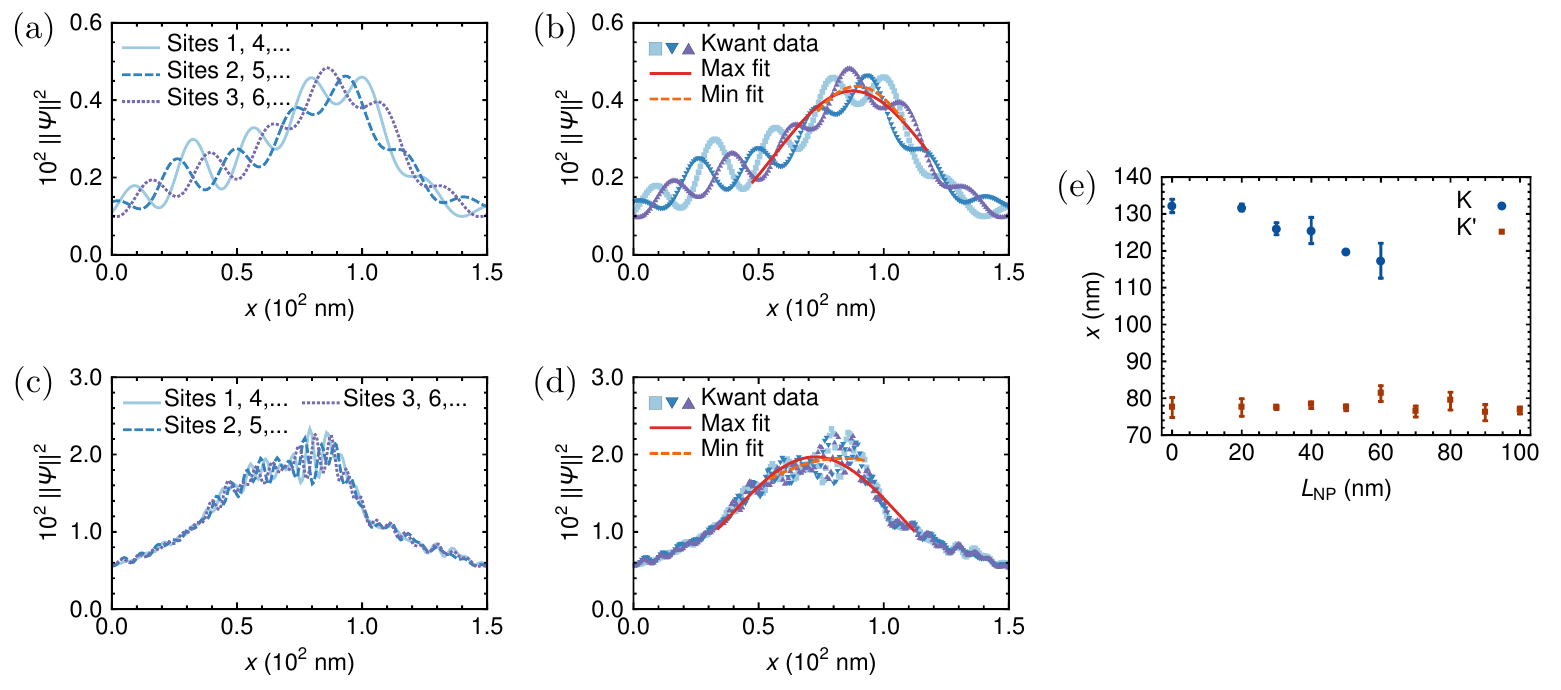}
  \caption{(a)--(d) Illustration of the fitting procedure (for $\theta=0$) that we use to determine the position of the maximum, as explained in the section on the details of the numerical simulations. 
    The simulation data show a clear three-site periodicity, which is particularly visible for low energies. This periodicity is most likely a consequence of the additional phase factor $\exp(2i \pi/3 \cdot m)$, with $m$ the site index, in the Bloch wavefunction. It arises from the fact that the $x$-coordinates of the points $K$ and $K'$ equal one third of the $x$-coordinate of a reciprocal lattice vector. In all four figures we show $| \Psi_{\text{av},\alpha} |^2$, where we have averaged over sublattices and have summed over all modes in the $K'$-valley. (a, b) The electron energy equals $E=0.1$~eV. The maxima of both Gaussian fits are close together, which means that the error bar is small. (c, d) The electron energy $E=0.4$~eV. The maxima are further apart, giving rise to a larger error bar. For both samples $W_i=40$~nm, $L_1=100$~nm and $U_0=2E$. (e) The dependence of the position of the maximum on the smoothness of the junction. Increasing the smoothness of the junction greatly lowers the intensity of both foci. For the $K'$-valley, the focus remains rather sharp and is recognizeable as such. For the $K$-valley, the focus becomes smeared, and the smearing becomes much stronger as the junction width increases. Beyond $L_{\text{NP}}=60$~nm it is hard to recognize a real focus for the $K$-valley, and we have therefore not plotted a maximum.}
  \label{fig:fits-real-space-and-smooth-barr}
  \end{figure*}

  We obtain the total wavefunction in valley $\alpha$ by adding the averaged wavefunctions of all of the modes in that valley, i.e. $| \Psi_{\text{av},\alpha} |^2 = \sum_n | \Psi_{\text{av},\alpha,n} |^2$, where we have suppressed the site index $i$. Subsequently, we determine the position of the focus on the line $y=0$. Since the maximum of the wavefunction is sensitive to the details of the sample and shows rather strong fluctuations, we do not consider it a suitable measure for the position of the focus. Instead, we consider a collection of data sets, and fit a Gaussian to each of them. To obtain this collection, we first compute the minimum and maximum of $|\Psi_{\text{av},\alpha}|^2$ and denote it by $I_{\text{min}}$ and $I_{\text{max}}$, respectively. Subsequently, we draw a line at $I_{\text{min}} + x(I_{\text{max}}-I_{\text{min}})$, where $x$ varies, and extract all points between its left-most and right-most intersections with $|\Psi_{\text{av},\alpha}|^2$. We form a collection of data sets by letting $x$ vary in steps of 0.01, typically between 0.3 and 0.7. After the fitting, we extract the position of the maximum of each of the Gaussians. By computing the average and standard deviation, we obtain a reliable value and an error for the position of the focus. Hence, the error bar that is shown in the figures does not reflect the width of the focus, but rather how well its position is defined. We refer to  Fig.~\ref{fig:fits-real-space-and-smooth-barr} for two examples of the fitting procedure.
  
\begin{figure}[!tbh]
  \includegraphics{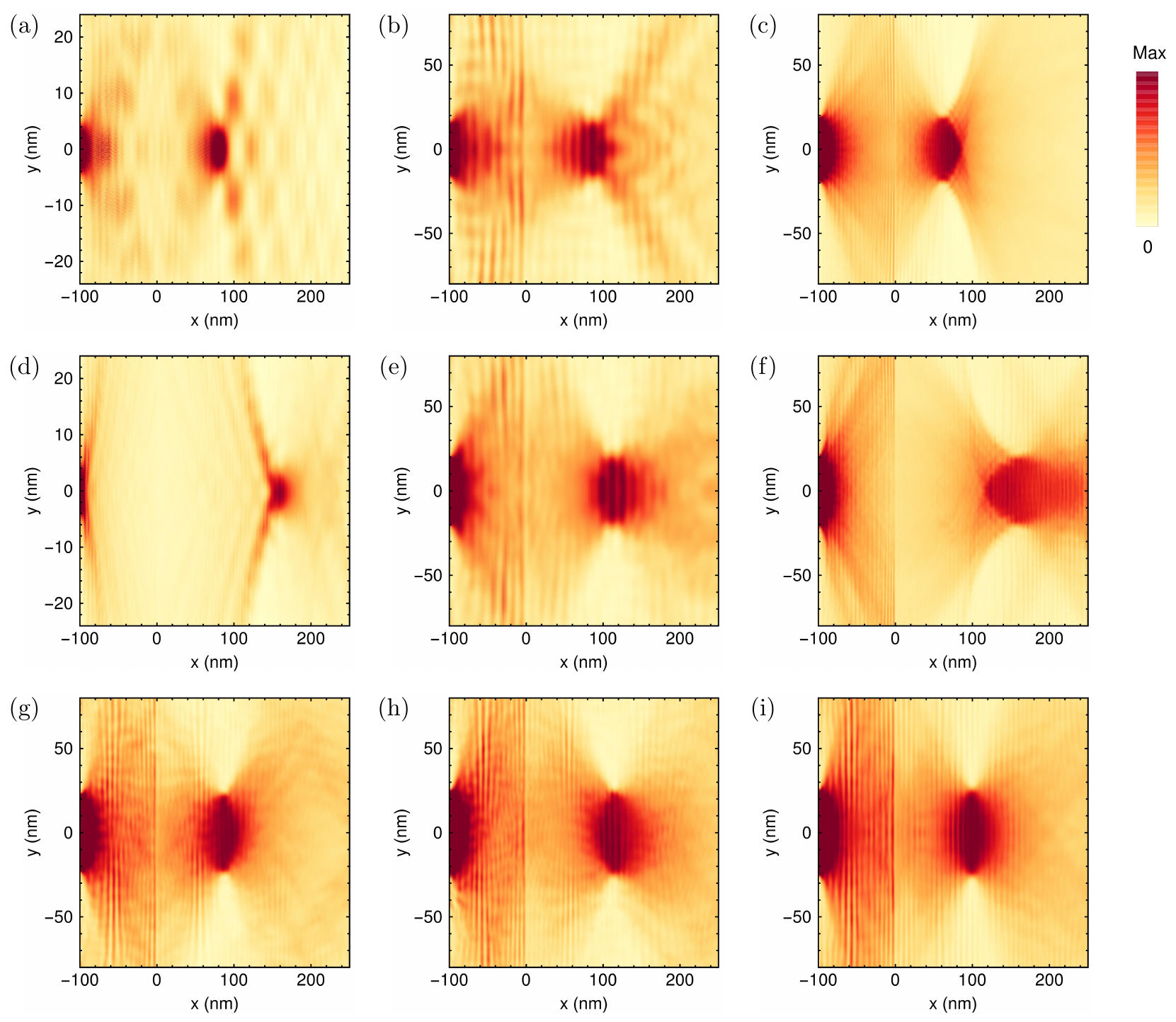}
  \caption{
    Results of the tight-binding simulations performed with the Kwant code for $L_1=100$~nm, $U_0=2E$. We plot the density $| \Psi_{\text{av},\alpha} |^2$, where we have averaged over sublattices and have summed over all lead modes for a given valley. The \emph{n-p} junction is at $x=0$. (a) Results for the $K'$-valley for a narrow lead of width $W_i = 16$~nm and electron energy $E=0.3$~eV. There is only one mode in the lead. (b) Results for a wider lead, $W_i =40$~nm, and $E=0.2$~eV. There are 3 modes in the $K'$-valley. (c) The results for electron energy $E=0.6$~eV, with $W_i =40$~nm. There are 11 modes in the $K'$-valley. (d) Results for the $K$-valley for a narrow lead of width $W_i=7.5$~nm and electron energy $E=0.6$~eV. There are two modes in the lead. (e)--(f) Results for the $K$-valley for the parameters used to generate the results in panels (b)--(c). The number of modes in the lead for the $K$-valley is one larger than the number of modes in the $K'$-valley. The dimensionless semiclassical parameter in the lead $\hbar v_F/(E W_i)$ equals (d) 0.142; (a) 0.133; (b, e) 0.080; (c, f) 0.027. When this parameter decreases, we clearly see that we go from a caustic to a sharp focussing spot, as predicted in Ref.~\cite{Reijnders17}.
    (g) A rotated sample with $\theta=19.1$ degrees. The lead with $W_i = 50$~nm and $E=0.4$~eV. There are 9 modes in the $K'$-valley. (h) The results for the $K$-valley for the same parameters as in (g). There are 10 modes in the lead. The tilting of the focus is opposite for the two valleys. (i) The results for a sample with $\theta=90$~degrees, corresponding to armchair edges along the $x$-axis. We clearly see a sharp focussing spot at $x=L_1$, in accordance with the classical trajectories in Fig.~\ref{fig:setup-trajectories}(g).
    The maximum of the color scale corresponds to: (a) 0.0096, (b, e) 0.0093, (c, f) 0.031, (d) 0.02, (g, h) 0.02, (i) 0.04.
  }
  \label{fig:extra-sims-realspace}
  \end{figure}

  For the rotated samples, we also extract the total wavefunction on the $x$-axis, and determine the position of the peak for this data. This is probably not the position of the true focus, since we know from a study of the trajectories that the cusp point does not lie at $y=0$, see Fig.~\ref{fig:setup-trajectories}.
  Another indication for this is given by the tilting of the focus in the plots of the wavefunction for the rotated samples, see Fig.~\ref{fig:extra-sims-realspace}. However, the differences are seen to be rather small, and we obtain a good indication of the position of the focus by considering the data on the $x$-axis.
  
  In most simulations, we consider an atomically sharp potential barrier. We have also done a few simulations for a smooth potential, see Fig.~\ref{fig:fits-real-space-and-smooth-barr}, which was implemented as
  \begin{equation}
    U(x) = \frac{U_0}{2} \left[ 1 + \tanh\left( \frac{3 x}{L_{\text{NP}}}  \right) \right] .
  \end{equation}
  We consider $L_{\text{NP}}$ to be a good measure for the length scale of the potential. Alternatively, one can compute the tangent to the potential $U(x)$ at $x=0$ and take the length scale of the potential to be the difference between the points $x_-$ and $x_+$, which are defined as the points where the tangent crosses zero and $U_0$, respectively. This leads to a width of $x_+-x_-=2 L_{\text{NP}}/3$.
  
  We obtain data for a varying potential by considering a collection of samples with different potentials. We typically divide the energy interval into 200 steps, and numerically compute the $S$-matrix for each value of the potential. In this case $W_c=W_i$, and $W=3 L_1+W_i$ for $L_1=L_2=100$~nm and $W=2 L_1+W_i$ when $L_1$ is larger. We obtain the tunneling coefficients for intravalley and intervalley scattering by adding the squared norms of the appropriate $S$-matrix elements and by subsequently dividing by the total number of modes in both valleys. We observe that intervalley scattering only mildly affects the results, see Fig.~\ref{fig:numerics-corrections}.

  \subsection{Details for the billiard model and probability distributions}

  In the billiard simulations, electrons are modelled as billiard balls~\cite{Beenakker89,Milovanovic13,Milovanovic15}. We consider $N$ electrons, with initial positions between $-W_i/2$ and $W_i/2$ that are drawn from a uniform distribution. Their initial transversal momenta are either drawn from a uniform distribution, or from a distribution that is uniform in the emission angle (see below). Subsequently, we compute their classical trajectories, and check whether the particles reach the collector lead, which has a width $W_c=W_i$. Furthermore, for each electron, we draw a random number between zero and one for each junction interface on its trajectory. When this number is smaller than the tunneling probability, the electron is transmitted by the interface; otherwise, it is reflected. We count the number $n_T$ of electrons that reach the collector lead, and define the tunneling probability as $T = n_T/N$. We observe a standard $1/\sqrt{N}$ convergence of $T$, roughly independent of the dispersion relation and the distribution of the transversal momenta. We use $N=250000$ in our simulations, for which the standard deviation of $T$ is approximately equal to 0.001.
  
  We first consider a uniform distribution of the initial transversal momenta. For a nonrelativistic (Schr\"odinger) particle in a box, the momenta are equally spaced, which gives rise to a uniform momentum distribution in the continuum limit. This model is therefore valid when $W_i \gg \lambda_{\text{el}}$, with $\lambda_{\text{el}}$ the (de Broglie) wavelength of the electrons, see also Ref.~\cite{Milovanovic13}.
  Although the momenta in a graphene ribbon are not precisely equally spaced~\cite{Brey06}, it was shown that a uniform momentum distribution gives correct predictions for graphene Hall barr experiments~\cite{Milovanovic13,Milovanovic15}. In the Dirac approximation, a uniform momentum distribution $f_{p_y}(p_y)$ between $-p_{y,\text{max}}/2$ and $p_{y,\text{max}}/2$ corresponds to a distribution $f_\phi(\phi) = \cos\phi/(2 \sin\phi_{\text{max}})$ of emission angles, where $p_{y,\text{max}}=E \sin\phi_{\text{max}}$. This can be easily verified using the formula
  \begin{equation}
    f_\phi(\phi) = f_{p_y}(p_y(\phi)) \frac{\text{d}p_y}{\text{d}\phi} ,
  \end{equation}
  and the relation $p_y=E\sin\phi$. We have verified that, within the Dirac approximation, both samplings indeed give the same result for the transmission. 

  We also consider a distribution that is uniform in the emission angle. This distribution, with a maximal emission angle of 45 degrees, was used in Ref.~\cite{Lee15}, where the authors used billiard simulations to model their experiments.
  Following their approach, we also set the maximal emission angle to 45 degrees.
  For the Dirac equation, the emission angles can be converted into momenta using the relation given above. When we include trigonal warping, this conversion is more complicated, since the emission angle is determined by the group velocity. In this case, we obtain the transversal momenta by numerically finding the root of Eq.~(\ref{eq:angle-py-trigwarp}).
  
\begin{figure*}[!ptbh]
  \includegraphics{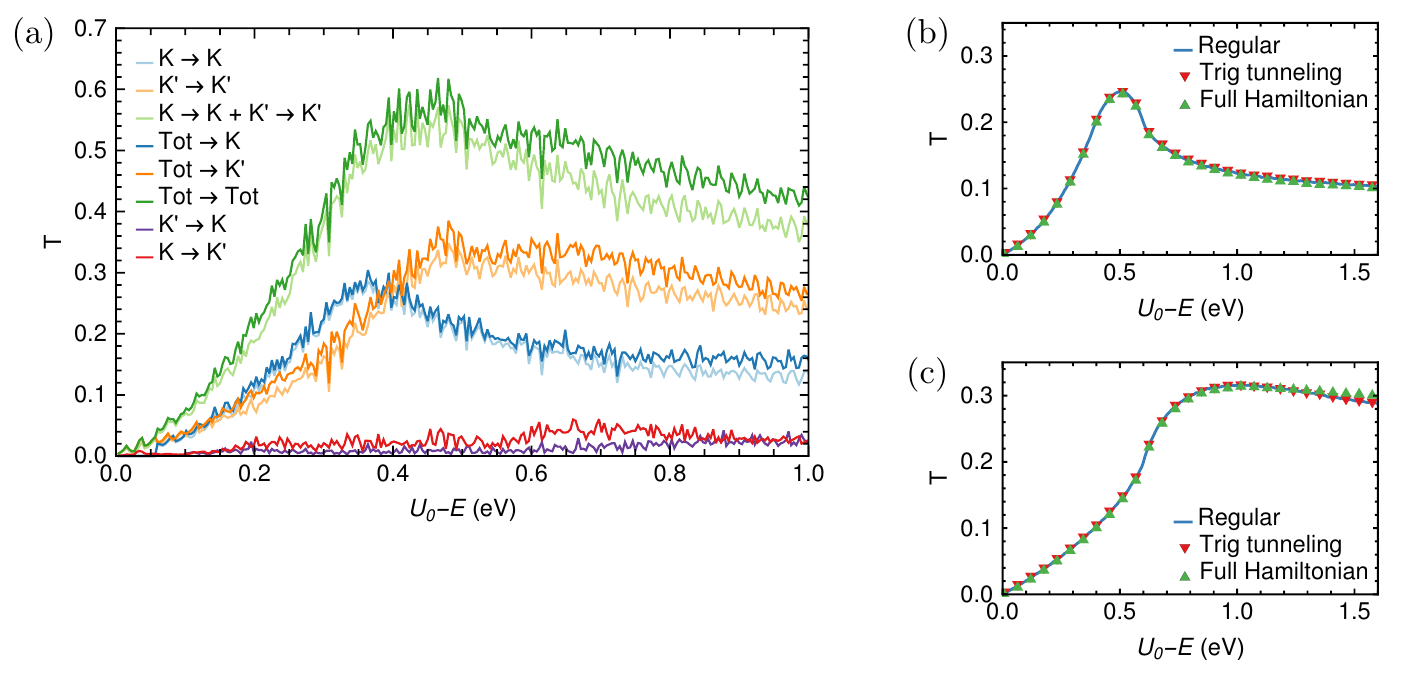}
  \caption{(a) Comparison of intravalley and intervalley scattering in a Kwant simulation for a sample with $W_i=W_c=50$~nm, $L_1=L_2=100$~nm and $E=0.4$~eV. We see that the effect of intervalley scattering on the transmission is small, its effect being slightly larger for larger energies. The essential features of the transmission into the valleys $K$ and $K'$ are preserved when intervalley scattering is added.
  (b, c) The effect of two corrections on the billiard model for a simulation with $W_i=W_c=50$~nm, $L_1=L_2=100$~nm and $E=0.6$~eV. Depicted are (b) the $K$-valley and (c) the $K'$-valley. The blue line shows the results for a simulation where the trajectories are computed using the trigonal warping Hamiltonian and the tunneling coefficient derived from the Dirac Hamiltonian is used. For the red downward triangles, the Dirac tunneling coefficient is replaced by the trigonal warping tunneling coefficient. We see that it does not essentially change the transmission. The green upward triangles show the results for a simulation where the trajectories are computed using the full nearest-neighbor tight-binding Hamiltonian~(\ref{eq:H-matrix-full}), while the Dirac tunneling coefficient is used. This only affects the transmission at very high energies. We see that its effect is stronger in the $K'$-valley, where the transmission at high energies is higher than in the $K$-valley.}
  \label{fig:numerics-corrections}
  \end{figure*}

  The classical trajectories of the electrons are determined from Eq.~(\ref{eq:trajectories}). The longitudinal momenta $p_x$ are computed by solving the equation $H_\alpha^\pm = E$ for $p_x$, given $p_y$. In the Dirac case this can be done analytically, but when we include trigonal warping we need to use numerical methods. 
  We also perform a calculation using the full nearest-neighbor tight-binding Hamiltonian, given by Eq.~(\ref{eq:H-matrix-full}), and find that the tunneling probabilities only differ for high energies, see Fig.~\ref{fig:numerics-corrections}.  
  Furthermore, we study the influence of a finite-size sample by allowing for a cutoff in the transversal direction. Taking out the electrons for which $|y(x=0)|>L_1+W_i/2$ in the simulations with uniformly distributed transversal momenta, we effectively limit the emission angle to 45 degrees, but find no significant influence on the results. We ascribe this to Klein tunneling, which collimates the electron beam, whence the largest contribution to the transmission comes from electrons that are emitted with small angles. We therefore plot the results without cutoff.
  Likewise, choosing maximal emission angles larger than 45 degrees in the model with uniformly distributed emission angles has a minimal effect on the position of the transmission maximum, although it does influence its value. Taking out electrons for which $|y(x=0)|>L_1+W_i/2$ at a maximal emission angle of 45 degrees also has very little influence.
  
  The tunneling coefficient through an \emph{n-p} junction is well-known for the Dirac equation~\cite{Katsnelson06,Reijnders17}. To obtain the tunneling coefficient when we include trigonal warping, we compute the eigenvectors of the matrix Hamiltonian, see Eqs.~(\ref{eq:H-matrix-trigwarp-K}) and~(\ref{eq:H-matrix-trigwarp-Kpr}), and normalize them to unit current. Subsequently, we match them at the barrier interface.
  We observe that replacing the trigonal warping tunneling coefficient by the Dirac tunneling coefficient does not essentially change the observed transmission, see Fig.~\ref{fig:numerics-corrections}. Since the latter can be computed much more efficiently than the former, we use the Dirac tunneling coefficient in our simulations. This means that trigonal warping is only incorporated into the trajectories.

  \subsection{Fitting the data for a varying potential}
  The data that we obtain when we vary the potential are rather smooth, both for the billiard simulations and for the simulations using Kwant. To further smoothen the data, we apply a Gaussian filter with radius $r$. For the Kwant data, this radius varies in integer steps between four and fourteen and for the billiard data it varies between two and seven. Subsequently, we compute the maximum for each of these smoothened data sets. We then average over the obtained maxima, and compute the standard deviation. For all cases, the standard deviation is smaller than the plot marker. To check our results, we also fit a Gaussian to a reasonable subset of the raw data and extract the position of its maximum. We find good agreement between the results of the two methods. The graphs show the results of the first procedure.
  
  To determine the width of the peak as a function of $L_1=L_2=L$, we once again smoothen the data by applying a Gaussian filter with radius $r$. For the Kwant data, this radius varies between four and fourteen. For the billiard data, it varies between two and four. We then compute the potentials $U_{0,\text{left}}$ and $U_{0,\text{right}}$, left and right of the transmission maximum, respectively, for which the transmission equals 80 percent of its maximal value. Averaging over the values obtained for different Gaussian filters and subtracting $U_{0,\text{left}}$ from $U_{0,\text{right}}$, we obtain our final result.

  \subsection{Data processing for simulations with a mass in the lead}
  In our simulations using the Kwant code, we include a mass term $m_{\text{lead}}$ in the lead by adding an on-site potential $m_{\text{lead}}$ for sites that belong to sublattice $A$, and $-m_{\text{lead}}$ for sites that belong to sublattice $B$. The obtained lead wavefunctions are in agreement with the results of the continuum model.
  
  The computations of the wavefunction in real space are performed in the same way as before. To quantify the asymmetry in the averaged wavefunction, we consider slices of the sample with a given $x$-coordinate, and subsequently compute the fraction $\eta_{b,x,\alpha}$ of the total intensity on the slice that is located on the sites with $y\leq 0$, that is,
  \begin{equation}
    \eta_{b,x,\alpha} = \bigg(\sum_{\substack{i \text{ where} \\ x_i = x \text{ and } y_i \leq 0}} | \Psi_{i,\text{av},\alpha} |^2 \bigg) \bigg/ \bigg(\sum_{\substack{i \text{ where} \\ x_i = x}} | \Psi_{i,\text{av},\alpha} |^2 \bigg) .
    \label{eq:def-fraction-bottom-real-space}
  \end{equation}
  Since the wavefunction at $x=-L_1$ is not symmetric in the line $y=0$, because the lead wavefunction of the incoming mode is not, we subtract the fraction $\eta_{b,-L_1,\alpha}$ at $x=-L_1$ to see the change in the intensity distribution.
  
  Furthermore, we consider the continuum evolution of the initial wavefunction outputted by the Kwant code. To this end, we extract the wavefunction from the Kwant simulation at the left-most point of the sample ($x=-L_1$). We split the sites into those belonging to sublattice $A$ and those belonging to sublattice $B$. Subsequently, we compute the evolution according to the continuum (Dirac) model, using Eq.~(42) from Ref.~\cite{Reijnders17}. We also compute the Fourier transform of the initial wavefunction using the equations from the same paper. Comparing these simulations with the simulations of the sample wavefunction using the Kwant code, we can establish which symmetry breaking effects can be explained by the continuum model, and which effects are due to trigonal warping.
  
  We also consider a second type of simulations to probe the asymmetry induced by a mass in the lead. Here, we consider two collector leads instead of one, with the first one, called the top collector lead, between $y=0$ and $y=W/2$, and the second one, called the bottom collector lead, between $y=-W/2$ and $y=0$. We then compute the $S$-matrix for scattering into both collector leads and extract the transmission probabilities $T_{\text{bc},n,\alpha}$, which shows which part of incoming mode $n$ in valley $\alpha$ is scattered into the bottom collector lead, and $T_{\text{tc},n,\alpha}$, which indicates the fraction scattered into the top collector lead.  
  Summing over all modes in a given valley, we obtain the total transmission probabilities $T_{\text{bc},\alpha}=\sum_n T_{\text{bc},n,\alpha}$ and $T_{\text{tc},\alpha}=\sum_n T_{\text{tc},n,\alpha}$.
  We then define the fraction that is transmitted into the lower lead as
  \begin{equation}
    \eta_{b,\text{out},\alpha} = \frac{T_{\text{bc},\alpha}}{T_{\text{bc},\alpha}+T_{\text{tc},\alpha}} .
  \end{equation}
  In order to obtain a measure for the asymmetry that is induced by the presence of the mass term, one should subtract the initial fraction of the intensity that is at negative $y$. Unfortunately, this is not a simple endeavor, since the total wavefunction in the lead is a sum of forward-moving modes and reflected modes and is not easily computed.
  Therefore, we confine ourselves to the forward-moving modes. For each mode, we extract the total intensity located on the sites with $y\leq 0$. Summing over all incoming modes and dividing by the total intensity, we obtain the fraction $\eta_{b,\text{in},\alpha}$ of the total intensity that is located on the sites with negative $y$. The procedure is the same as in Eq.~(\ref{eq:def-fraction-bottom-real-space}), the only difference being that we do not average over sublattices this time.
  We obtain the dependence of the asymmetry $\eta_{b,\text{out},\alpha}-\eta_{b,\text{in},\alpha}$ on $L_2$ by computing the $S$-matrix for a collection of samples with varying $L_2$ and fixed energy, mass and potential. We have checked that the measure $\eta_{b,\text{out},\alpha}-\eta_{b,\text{in},\alpha}$ is roughly independent of the width of the sample and is roughly equal to zero for zero mass (see Fig.~\ref{fig:mass-influence-main}), making it a suitable measure for the asymmetry. Furthermore, it evolves smoothly with changing width of the lead.
  
  Finally, we compute the dependence of the transmission on the potential strength for samples with $W_c=W_i$ in the presence of a mass $m_{\text{lead}}$ in the lead. The computations using the Kwant code are done in the same way as before, the only difference being the mass in the lead. Since the billiard model does not include information on the initial wavefunction, the results for this model are not affected by the presence of a mass term in the lead.

  \clearpage

  \section{Additional simulation results}

  \subsection{Kwant simulations in real space}
  
  \clearpage

  \subsection{Determination of the maximum}

  \clearpage

  \subsection{Higher order effects in the numerical simulations}
  
  \clearpage

  \subsection{Transmission as a function of potential strength}

  \begin{figure*}[!ptbh]
  \includegraphics{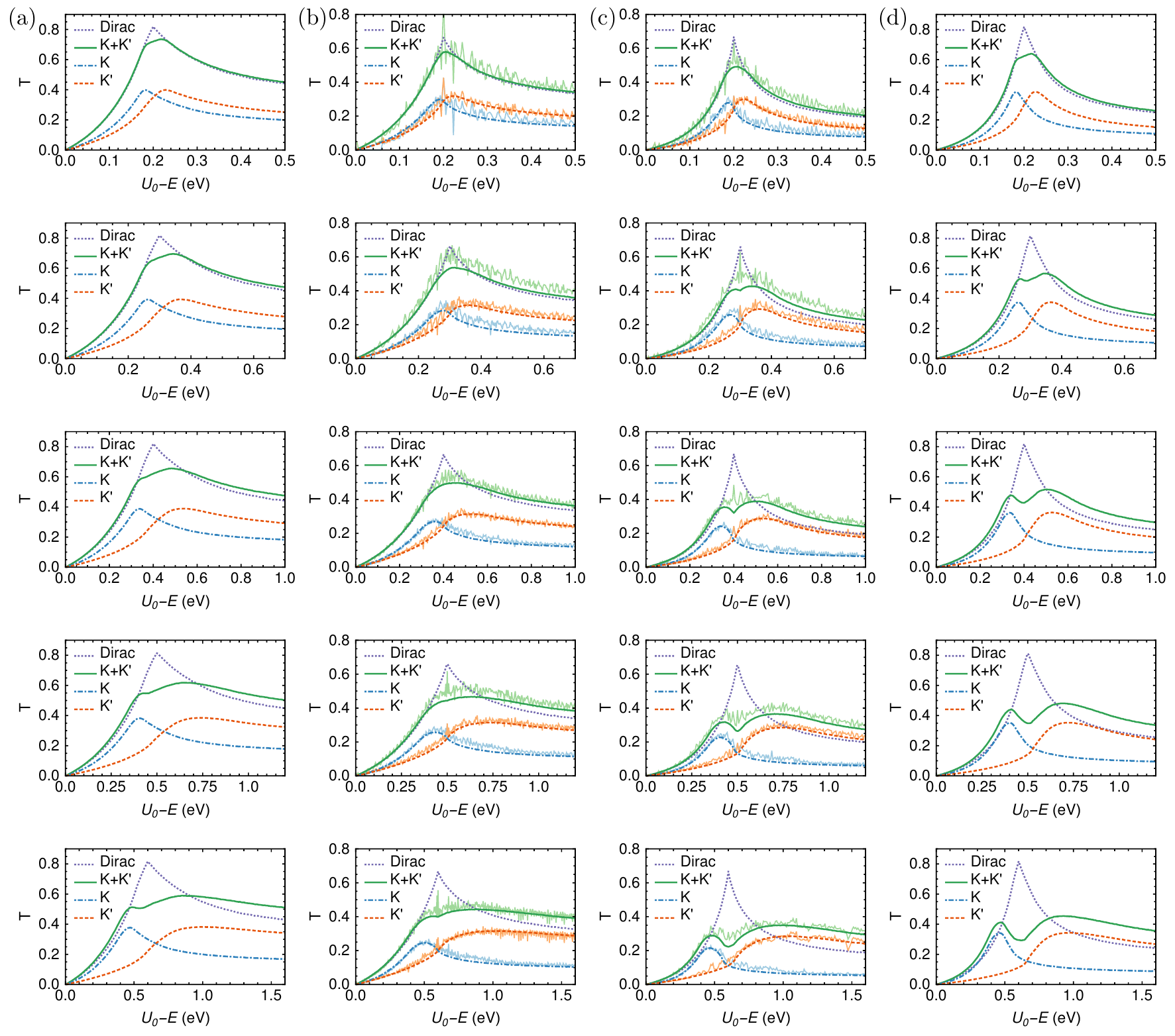}
  \caption{Transmission through an \emph{n-p} junction as a function of potential strength for various setups. The rows correspond to different energies, namely $E=0.2$~eV (top row), $E=0.3$~eV, $E=0.4$~eV, $E=0.5$~eV and $E=0.6$~eV (bottom row). (a) In the left column, we consider a billiard model where the emission angles are uniformly distributed, with maximal emission angle $\pi/4$. The sample length $L_1=L_2=100$~nm. We see that the peaks for both valleys are approximately equal in height. (b) In the second column, we consider the same sample length as in the left column. We compare the results that are obtained from Kwant simulations (light colors) with the results from our billiard model where the initial transversal momenta are uniformly distributed (dark colors). We see that the peaks for the two valleys are no longer equal in height and that the two simulations show very good agreement at all energies. Note that the Kwant data do not match a billiard model with uniformly distributed emission angles. (c) In the third column, we compare the same two quantities as in (b) for $L_1=L_2=200$~nm. (d) In the right column, we show the results from the billiard model considered in (a) for $L_1=L_2=200$~nm. For all samples $W_i=W_c=50$~nm.
  Comparing the different rows, one clearly sees that the splitting between the peaks for $K$ and $K'$ increases when the energy increases. Comparing the different columns, one sees that the peak splitting becomes more pronounced at larger length scales.
  }
  \label{fig:T-U-dep-trends}
  \end{figure*}
  
  \clearpage
  
  \begin{figure*}[!ptbh]
  \includegraphics{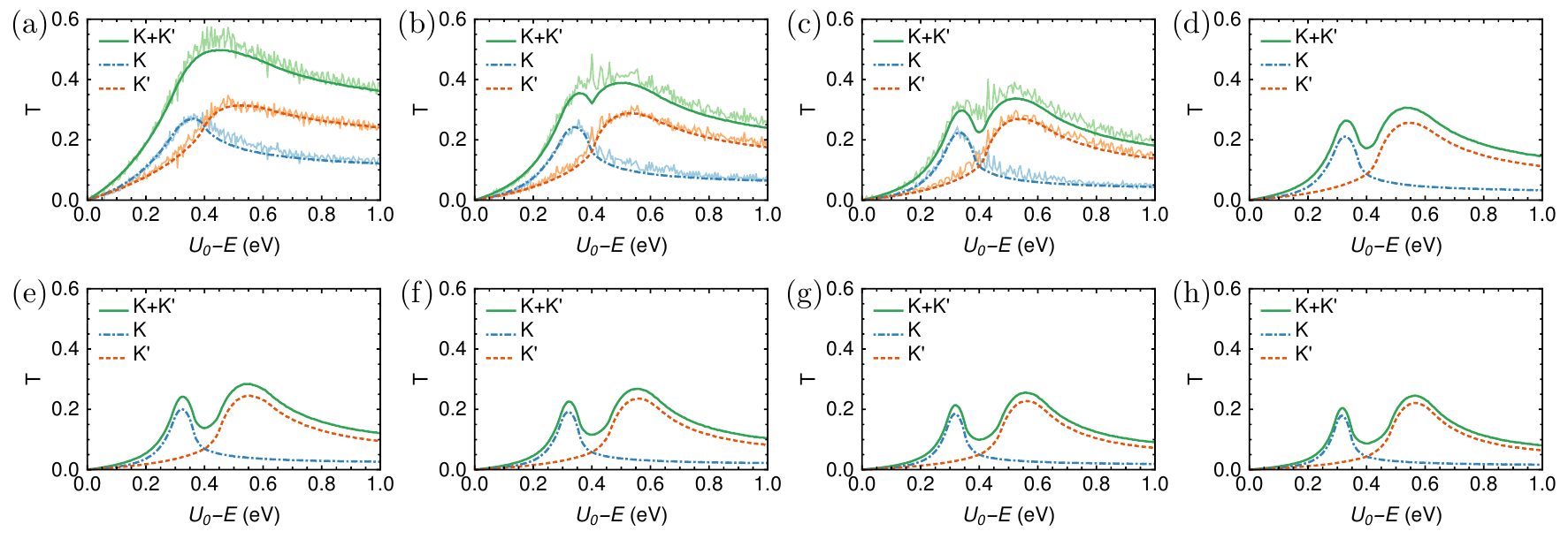}
  \caption{Transmission through an \emph{n-p} junction as a function of potential strength for various length scales $L_1=L_2=L$. We consider a sample with $W_i=W_c=50$~nm and an electron energy $E=0.4$~eV. We show the results of the billiard model with uniformly distributed transversal momenta (dark colors) for (a) $L=100$~nm, (b) $L=200$~nm, (c) $L=300$~nm, (d) $L=400$~nm, (e) $L=500$~nm, (f) $L=600$~nm, (g) $L=700$~nm, (h) $L=800$~nm. For panels (a)--(c), we also show the results of the Kwant simulations (light colors). The potential $U_{0,\text{max},\alpha}$, at which the transmission is maximal, only changes slightly when we increase $L$. However, increasing the length scale reduces the peak width (and also the transmission), which makes the splitting more pronounced.
  }
  \label{fig:T-U-Ldep}
  \end{figure*}

  We note that in the simulations of the billiard model with uniformly distributed transversal momenta, reported in Fig.~\ref{fig:T-U-dep-trends}(b) and~(c) and in Fig.~\ref{fig:T-U-Ldep}, the transmission probabilities for the valleys $K$ and $K'$ are equal at $U_0=2E$. To explain this, we first note that the sample that we are considering is symmetric in the junction interface ($x=0$), as $L_1=L_2=L$ and $W_c=W_i$. Second, we note that for $\theta=0$ and at $U_0=2E$ there is a special relation between electron and hole modes in the two valleys. We have $p_{x,e,K}(p_y)=-p_{x,h,K'}(p_y)$ and $p_{x,h,K}(p_y)=-p_{x,e,K'}(p_y)$. To convince oneself of these relations, one can visualize the energy contours of the dispersion relation~(\ref{eq:E-class-trigwarp-dimless-noAngle}) and invoke electron-hole symmetry.
  
  Alternatively, one can also show the above relations explicitly, though the derivation is slightly cumbersome. Consider an electron in valley $K$, with energy $E$ and transversal momentum $p_y$. Let $p_{x,e,K}(p_y)$ be the associated longitudinal momentum, for which the relation
  $ E = | \mathbf{p} | - \mu | \mathbf{p} |^2 \cos (3 \phi_{\mathbf{p}} ) $
  holds, see Eq.~(\ref{eq:H-class-trigwarp-dimless}). We now note that $-\cos(3 \phi_{\mathbf{p}})=\cos(3 [\pi-\phi_{\mathbf{p}}])$. Multiplying both sides by minus one and adding $2E$, we obtain the relation 
  $ E = -| \mathbf{p} | - \mu | \mathbf{p} |^2 \cos (3 [\pi-\phi_{\mathbf{p}}] ) + 2E$. Looking at Eq.~(\ref{eq:H-class-trigwarp-dimless}), we see that this is exactly the dispersion relation for a hole with momentum $-p_{x,e,K}(p_y)$ in valley $K'$ at $U_0=2E$. Therefore, we arrive at the relation $p_{x,e,K}(p_y)=-p_{x,h,K'}(p_y)$. The second relation can be establish analogously.
  
  Let us now consider an electron in the $K$-valley, which is emitted from the point $y_i$ in the injector lead with transversal momentum $p_{y,s}$ and longitudinal momentum $p_{x,e,K}(p_{y,s})$. This electron hits the junction interface at the point $y_e$, after which its longitudinal momentum becomes $p_{x,h,K}(p_{y,s})$. We first assume that it is transmitted to the point $y_c$ in the collector lead. Let us now also consider an electron in the $K'$-valley, which is emitted from the point $y_c$ in the injector lead with transversal momentum $p_{y,s}$. Then, because of the above relations, its longitudinal momentum equals $p_{x,e,K'}(p_{y,s})=-p_{x,h,K}(p_{y,s})$. Since $L_1=L_2$, this electron hits the junction interface at the same coordinate $y_e$, after which its longitudinal momentum becomes $p_{x,h,K'}(p_{y,s})=-p_{x,e,K}(p_{y,s})$. Because $L_1=L_2$, this particle subsequently reaches the collector lead at the point $y_i$ and is therefore transmitted. This means that we have established a one-to-one relationship between trajectories in the $K$-valley that are transmitted and trajectories in the $K'$-valley that are transmitted. When the electron in the $K$-valley that we considered does not reach the collector lead, the above procedure gives us an invalid trajectory in the $K'$-valley, since this new trajectory does not start from a point in the lead.
  
  We also note that, since we have used the Dirac tunneling probability in our simulations, the tunneling probability at the barrier interface does not depend on the valley. Then, because of the one-to-one relationship between transmitted trajectories in the two valleys, the amount of electrons transmitted in valley $K$ equals the amount of electrons transmitted in valley $K'$. Of course, this supposes that our simulations have converged, i.e. that we have sampled a sufficiently large number of initial positions and initial transversal momenta.
  
  With the above argument, we have shown that, at $U_0=2E$, the transmission probabilities for the valleys $K$ and $K'$ are equal within the framework of a billiard model with uniformly distributed transversal momenta. In principle, this argument should also hold for a billiard model with uniformly distributed emission angles. However, in our simulations, we have limited the maximal emission angle to $\pi/4$. This breaks the above argument, as it is not certain that the two related trajectories in the valleys $K$ and $K'$ both have emission angles smaller than $\pi/4$. Indeed, in Fig.~\ref{fig:T-U-dep-trends}(a) and~(d), we observe that for a billiard model with uniformly distributed emission angles the two transmission probabilities are not equal at $U_0=2E$.

  \vspace*{2cm}
  
  \begin{figure*}[!bh]
  \includegraphics{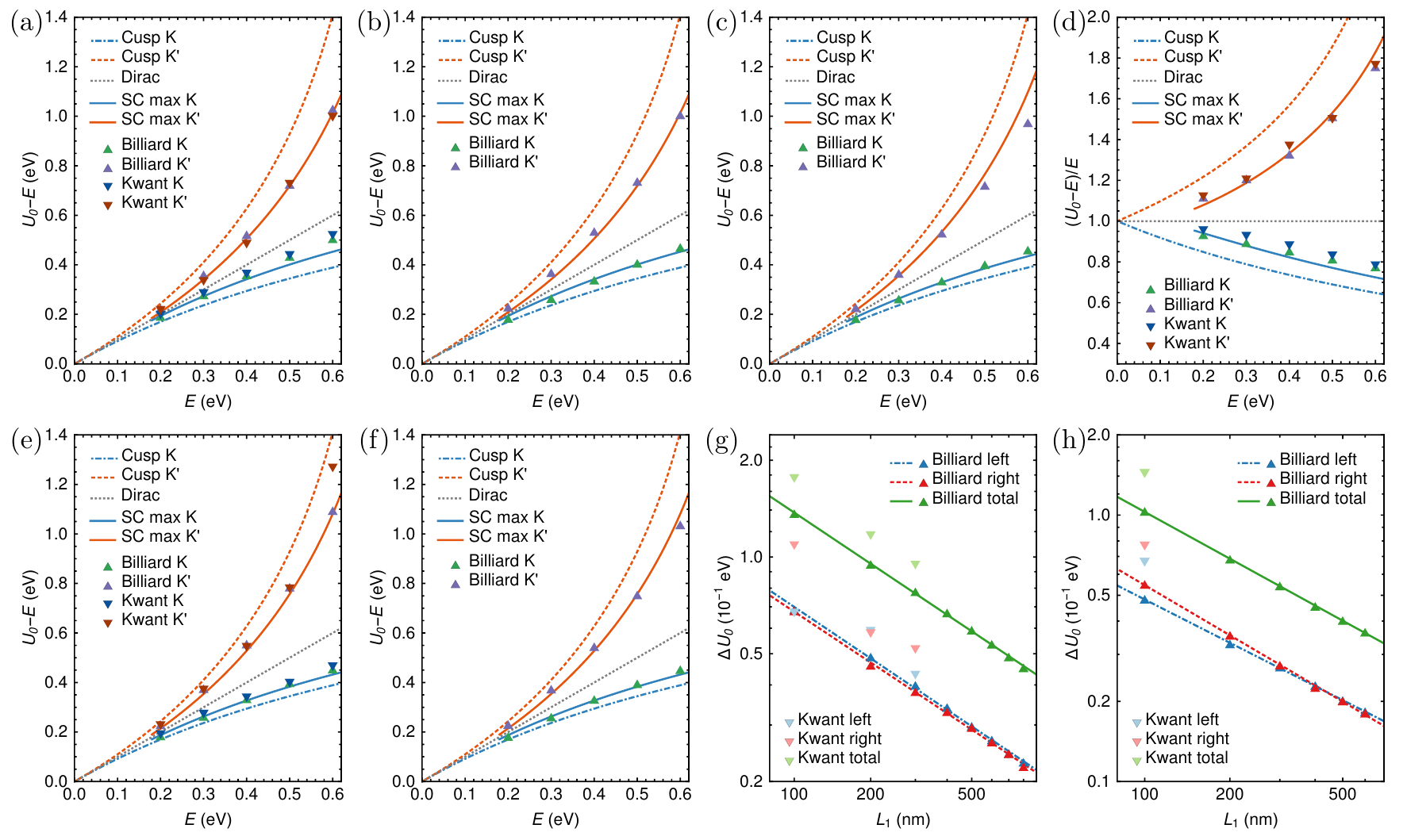}
  \caption{(a)--(f) Maximum of the simulated intensity as a function of energy. We also plot the potentials at which $x_{\text{cusp},\alpha}$ and the semiclassical (SC) $x_{\text{max},\alpha}$ equal $L_2$. (a) \emph{n-p} junction with $L_1=L_2=L=100$~nm. The billiard model has uniformly distributed transversal momenta. (b) \emph{n-p} junction with $L=100$~nm. The billiard model has uniformly distributed emission angles. (c) Same as (a) for $L=200$~nm. (d) Same as (b) for $L=200$~nm. As have we already shown the results for $U_0-E$ for this case in Fig.~\ref{fig:T-U-dep}(g), we now plot $(U_0-E)/E$ on the vertical axis. (e) The results for an \emph{n-p-n} junction where the electrons regions have length $L_1=L_3=L=100$~nm and the hole region has length $L_2=2L=200$~nm. We show a billiard model with uniformly distributed transversal momenta. (f) We consider the same sample as in (e), but this time we consider a billiard model with uniformly distributed emission angles. (g)--(h) The dependence of the peak width on the length scale of the sample for $E=0.4$~eV, for both the Kwant simulations and for a billiard model with uniformly distributed transversal momenta. We plot $\Delta U_{0,\text{left}} = U_{0,\text{max}}-U_{0,\text{left}}$, $\Delta U_{0,\text{right}} = U_{0,\text{right}}-U_{0,\text{max}}$ and $\Delta U_0 = U_{0,\text{right}}-U_{0,\text{left}}$, where $U_{0,\text{left}}$ and $U_{0,\text{right}}$ were defined in the section on how we fit the data for a varying potential. (g) The results for the $K$-valley for an \emph{n-p} junction with $L_1=L_2=L$. For the billiard model, all three differences scale as $L^{-\alpha}$, with $\alpha$ close to 0.5. The Kwant simulations show a larger peak width, with scaling behavior that is somewhat less clear. For the $K'$-valley (not shown), the scaling behavior of the billiard results is much more asymmetric, with both $\Delta U_{0,\text{left}}$ and $\Delta U_{0,\text{right}}$ not scaling as $L^{-1/2}$, but rather as $L^{-0.32}$ and $L^{-0.64}$, respectively. However, the difference $\Delta U_0$ does show scaling with an exponent close to $-0.5$. (h) The results for the $K$-valley for an \emph{n-p-n} junction with length scales $L_1=\tfrac{1}{2} L_2= L_3=L$. The scaling exponent $\alpha$ is slightly higher than $0.5$, with $\Delta U_0$ scaling as $L^{-0.58}$. For the $K'$-valley (not shown), the peak once again scales asymmetrically, though the asymmetry is smaller than for the \emph{n-p} junction. The total width $\Delta U_0$ scales as $L^{-0.59}$ for the $K'$-valley. For all panels (a)--(h) we have $W_i=W_c=50$~nm.
  }
  \end{figure*}

  \clearpage
  
  \subsection{Numerical results for the higher order singularity}
  
  \clearpage
  
  \subsection{Breaking of sublattice symmetry in the lead}
  
  \begin{figure*}[!tbh]
  \includegraphics{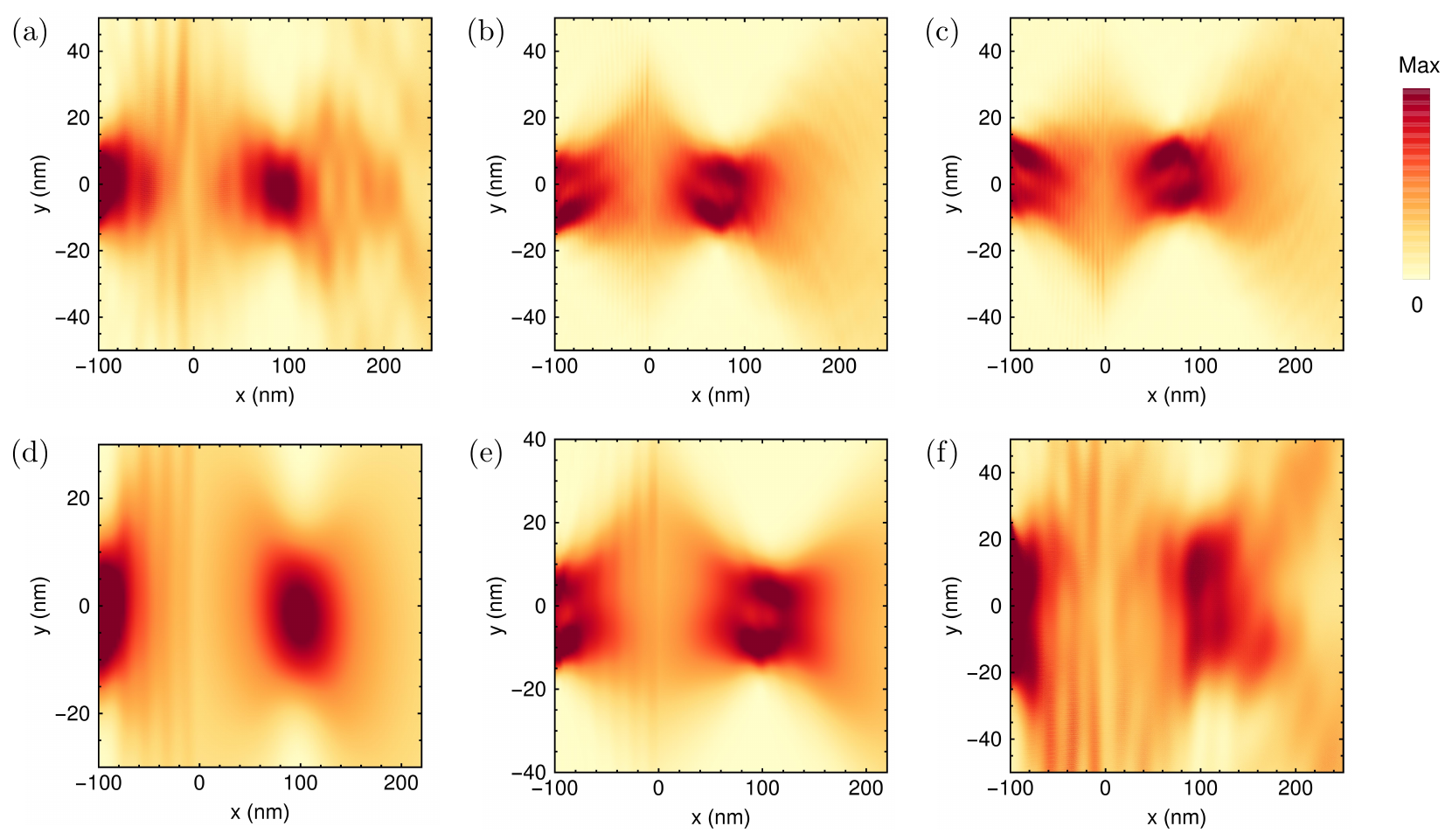}
  \caption{
    (a)--(c), (f) Results of the Kwant simulations when we include a constant mass $m_{\text{lead}}$ in the lead. We plot the density $| \Psi_{\text{av},\alpha} |^2$, where we have averaged over sublattices and have summed over all lead modes for a given valley. All figures are made using samples with the \emph{n-p} junction at $x=0$, with length scale $L_1=100$~nm and potential strength $U_0=2 E$.
    (a) Results for the $K'$-valley for $E=0.1$~eV, $m_{\text{lead}}=0.075$~eV and $W_i=50$~nm. There is only one mode in the lead and the focus is tilted downward.
    (b) Results for the $K'$-valley for $E=0.4$~eV, $m_{\text{lead}}=0.375$~eV, $W_i=40$~nm. There are two modes in the lead and the focus is tilted downward. (c) Results for $m_{\text{lead}}=-0.375$~eV, with the other parameters the same as in (b). We see that flipping the mass reflects the wavefunction in the $x$-axis.
    (f) Results for the $K$-valley for $E=0.1$~eV, $m_{\text{lead}}=0.075$~eV and $W_i=50$~nm. Summing over the two modes in the lead, an upward tilting is visible, though the focus is not very clear. 
    (d)--(e) Continuum evaluation of the initial condition taken from the Kwant simulation, for the $K'$-valley. The wavefunction at $x=-L_1$ was extracted from Kwant and this wavefunction was subsequently evolved using the continuum (Dirac) formulas from Ref.~\cite{Reijnders17}. In panel~(d) the parameters correspond to those in panel~(a) of this figure. In panel~(e) the parameters correspond to those in panel~(b) of this figure and Fig.~\ref{fig:mass-influence-main}(a) in the main text. One clearly sees that the tilting of the main focus is reproduced by the continuum evaluation, showing that it comes from the symmmetry breaking in the initial condition. However, since trigonal warping is not included into the continuum (Dirac) equations, the focus is located at $x=L_1$ in this approach.
    The maximum of the color scale corresponds to (a) 0.0044, (b, c) 0.0041, (d) 0.0064, (e) 0.007.
  }
  \end{figure*}

  \begin{figure*}[t]
  \includegraphics{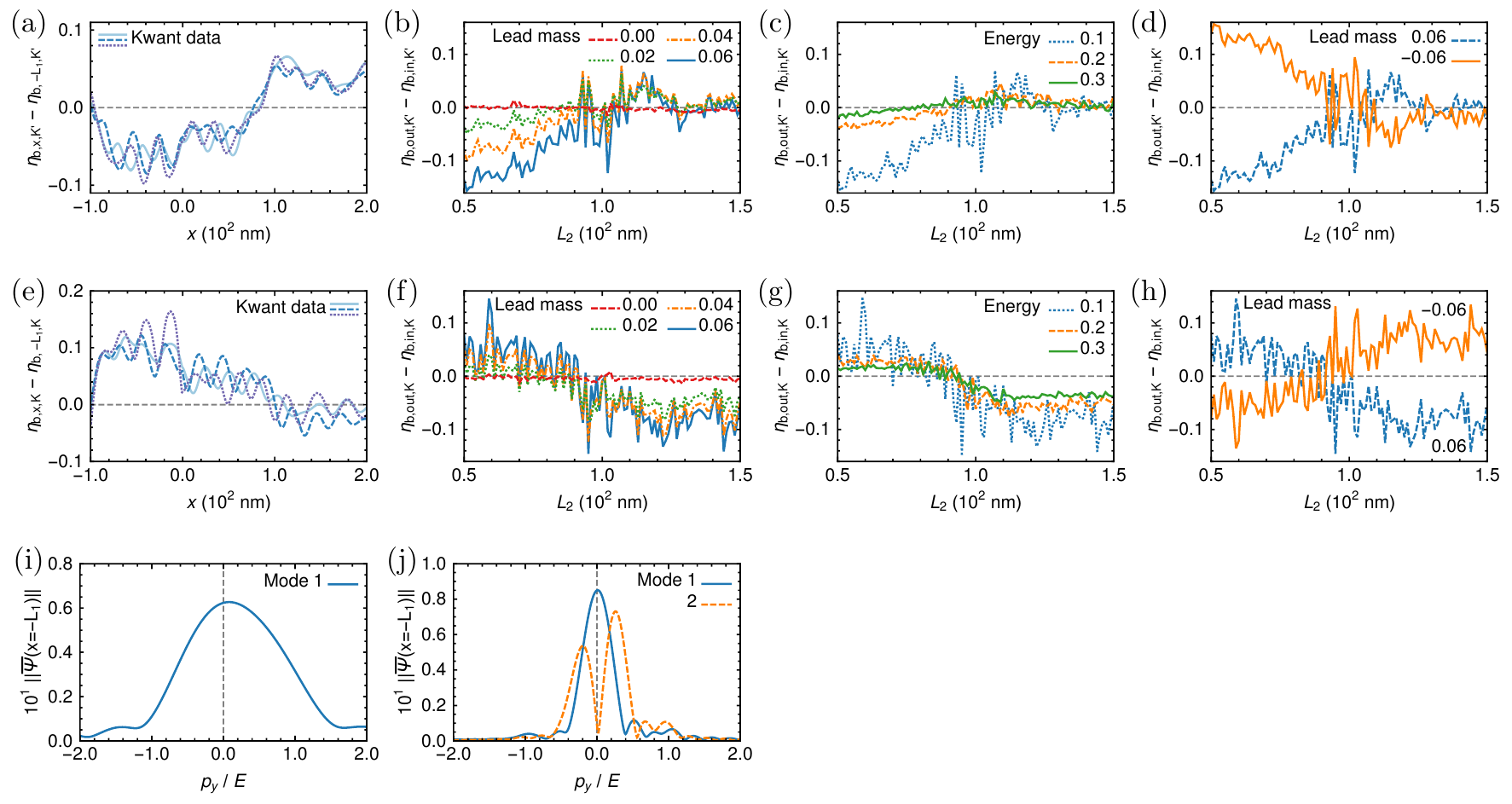}
  \caption{(a) Size of the asymmetry $\eta_{b,x,K'}-\eta_{b,-L_1,K'}$ as a function of position, computed according to the procedure outlined in the section on data processing. As in Fig.~\ref{fig:fits-real-space-and-smooth-barr}, we observe a three-site periodicity. This plot was constructed for the $K'$-valley, with $E=0.1$~eV and $m_{\text{lead}}=0.06$~eV. 
  (b) Size of the asymmetry $\eta_{b,\text{out},K'}-\eta_{b,\text{in},K'}$  as a function of $L_2$, for various masses (in eV). Simulations for the $K'$-valley, with $E=0.1$~eV. 
  (c) Size of the asymmetry as a function of sample length $L_2$, for various energies (in eV). Data were taken for the $K'$-valley, with $m_{\text{lead}}=0.6 E$.
  (d) Size of the asymmetry as a function of $L_2$, for positive and negative lead masses (in eV). Data for the $K'$-valley, with $E=0.1$~eV.
  For all four samples $U_0=2E$, $L_1=100$~nm and $W_i=50$~nm. Panels (e)--(h) depict the results for the $K$-valley for the parameters used in panels (a)--(d).
  (i)--(j) Fourier transform of the Kwant wavefunction at $x=-L_1$. (i) When $W_i=50$~nm and the electron energy $E=0.1$~eV, with $m_{\text{lead}}=0.06$~eV, the lead only accomodates one mode. Its Fourier transform shows a large asymmetry. (j) When the energy $E=0.4$~eV, with $m_{\text{lead}}=0.375$~eV and $W_i=40$~nm, there are two modes in the lead. The second mode shows a stronger asymmetry than the first one.
  }
  \label{fig:asymm-mass}
  \end{figure*}

  \begin{figure*}[t]
  \includegraphics{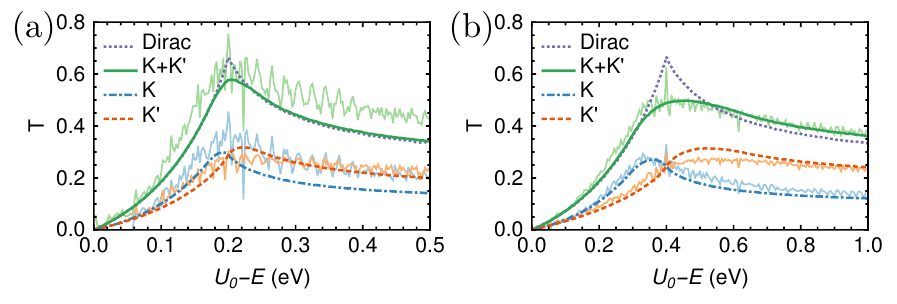}
  \caption{
  Transmission through an \emph{n-p} junction as a function of the potential strength $U_0$ when there is a constant mass $m_{\text{lead}}$ in the lead. (a) $E=0.2$~eV, (b) $E=0.4$~eV. For both cases $m_{\text{lead}}=0.75 E$, $L_1=L_2=100$~nm and $W_i=W_c=50$~nm. When we invert the signs of the lead masses, we obtain exactly the same figures.
  }
  \label{fig:ft-modes-lead}
  \end{figure*}

\end{document}